\documentclass[aps,twocolumn,showpacs,amsmath,amssymb,superscriptaddress,longbibliography,prl]{revtex4-1}
\usepackage[colorlinks=true,citecolor=blue,linkcolor=blue,hypertexnames=false]{hyperref}
\usepackage{amsmath}
\usepackage{dsfont}
\usepackage{hyperref}
\usepackage{textcomp}
\usepackage{tabularx}
\usepackage{graphicx}
\usepackage{soul}

\newcommand{\bk}{{\bf k}}

\newcommand{\bq}{{\bf q}}

\newcommand{\bj}{{\bf j}}

\newcommand{\bA}{{\bf A}}

\def \g{{\gamma}}

\def \bs{\boldsymbol}
\newcommand{\tl}{{\tilde \lambda}}

\newcommand{\w}{\omega}

\newcommand{\W}{\Omega}

\newcommand{\D}{\Delta}

\newcommand{\bp}{{\bf p}}
\newcommand{\vf}{v_F}

\def \be{\begin{equation}}
\def \ee{\end{equation}}
\def \bea{\begin{eqnarray}}
\def \eea{\end{eqnarray}}
\def \nn{\nonumber}

\newcommand{\old}[1]{\iffalse #1 \fi}
\begin{document}

\newcommand\rmf[1]{\textcolor{magenta}{#1}}

\title{Synergetic ferroelectricity and superconductivity in zero-density Dirac semimetals near quantum criticality}

\author{Vladyslav Kozii}
\affiliation{Department of Physics, Carnegie Mellon University, Pittsburgh, Pennsylvania 15213, USA}
\affiliation{Department of Physics, University of California, Berkeley, California 94720, USA}
\affiliation{Materials Sciences Division, Lawrence Berkeley National Laboratory, Berkeley, California 94720, USA}

\author{Avraham Klein}
\affiliation{Physics Department, Ariel University, Ariel 40700, Israel}
\affiliation{School of Physics and Astronomy, University of Minnesota, Minneapolis, Minnesota 55455, USA}

\author{Rafael M. Fernandes}
\affiliation{School of Physics and Astronomy, University of Minnesota, Minneapolis, Minnesota 55455, USA}

\author{Jonathan Ruhman}
\affiliation{Department of Physics, Bar-Ilan University, 52900, Ramat Gan, Israel}
\affiliation{Center for Quantum Entanglement Science and Technology, Bar-Ilan University, 52900, Ramat Gan, Israel}

\date{\today}

\begin{abstract}
We study superconductivity in a three-dimensional zero-density Dirac semimetal in proximity to a ferroelectric quantum critical point. We find that the interplay of criticality, inversion-symmetry breaking, and Dirac dispersion gives rise to a robust superconducting state  at the charge-neutrality point, where no Fermi surface is present. Using Eliashberg theory, we show that the ferroelectric quantum critical point is unstable against the formation of a ferroelectric density wave (FDW), whose fluctuations, in turn, lead to a first-order superconducting transition. Surprisingly, long-range superconducting and FDW orders are found to cooperate with each other, in contrast to the more usual scenario of phase competition. Therefore, we suggest that driving charge neutral Dirac materials, e.g., Pb$_x$Sn$_{1-x}$Te, through a ferroelectric quantum critical point may lead to superconductivity intertwined with FDW order.
\end{abstract}

\maketitle

{\it Introduction.---}Superconductivity (SC) is observed in numerous doped materials with extremely low density of free charge carriers. Examples include doped SrTiO$_3$~\cite{SchooleyHoslerCohen64,BehniaPRL14,Littlewoodetal2019},
Sr-doped Bi$_2$Se$_3$~\cite{SrBiSe2015}, YPtBi~\cite{Butchetal11}, Tl-doped PbTe~\cite{TlPbTe06}, and elemental bismuth~\cite{Prakash17}. The observation of low-density SC is surprising for two main reasons~\cite{KoziiBiRuhman2019,GastiasoroRuhmanFernandes2020}. First, because the Fermi  and the Debye energies are similar in magnitude and therefore the Coulomb repulsion is na{\"i}vely unscreened. 
Second, the density of states of a three-dimensional Fermi liquid, which affects the transition temperature drastically in conventional superconductors, is orders of magnitude smaller than in standard superconducting alloys. Therefore, finding microscopic models with superconducting instabilities at arbitrarily low density is an outstanding challenge.

A necessary ingredient for realizing low-density SC is a sufficiently long-ranged attractive interaction~\cite{GurevichLarkinFirsov62,GastiasoroChubukovFernandes2019,Gastiasoro2020}. Such interaction can be provided by the fluctuations of a bosonic order parameter in the vicinity of a quantum critical point (QCP)~\cite{ChubukovSchmalian2005,Ledereretal2015,Metlitskietal2015,Wangetal2016}. In particular, it was proposed that superconductivity in  some low-density SCs, such as doped SrTiO$_3$ and Pb$_{x}$Sn$_{1-x}$Te, originates from the proximity to a ferroelectric (FE) QCP~\cite{rowley2014ferroelectric,Balatsky2015,WolfleBalatsky2018,Kedem2018,KanasugiYanase2018,GamboaVerri2018,rischau2017ferroelectric,Tomiokaetal2019,enderlein2020superconductivity,Gastiasoro2020}. 
%Importantly, PbTe and SnTe naturally lie close to a ferroelectric transition . 
%The FE phase is characterized by a finite lattice electric dipole moment, corresponding to an optical phonon displacement. 
Indeed, experiments find that the electronic properties of SrTiO$_3$ are strongly influenced by the transition~\cite{Stuckyetal2016,sakai2016critical,rischau2017ferroelectric,Herrera_2019,engelmayer2019ferroelectric,wang2019charge,enderlein2020superconductivity,Sapkota2020} and possibly also of Pb$_{x}$Sn$_{1-x}$Te~\cite{Parfenev2001}.

The soft modes near the critical point in polar crystals are transverse. According to the standard description of the electron-phonon interaction, transverse phonons are decoupled from the itinerant electron's density ~\cite{ruhman2019comment,WolfleBalatsky2018}.  Several mechanisms have been proposed to circumvent this issue, such as coupling to gapped longitudinal modes~\cite{WolfleBalatsky2018,enderlein2020superconductivity}, two transverse phonon exchange~\cite{Ngai,vandermarelpossible,kumarquasiparticle,volkov2021superconductivity,kiseliov2021theory}, and a linear vector coupling in multi-orbital systems~\cite{FalkoNarozhny2006,fu2015parity,KoziiFu2015,RuhmanKoziiFu2017,WuMartin2017,KoziiBiRuhman2019,Gastiasoro2020,volkov2020multiband,Gastiasoro2021}. A renormalization group (RG) analysis  found the vector coupling is marginally relevant in  Dirac semimetals~\cite{KoziiBiRuhman2019}. Based on this observation, it was shown that FE critical fluctuations are a promising pairing mechanism for low-density SC. However, close enough to the QCP these arguments break down since the system flows to strong coupling.

In this paper, we investigate the fate of the  clean ferroelectric QCP in 3D polar Dirac materials at zero density using both Eliashberg and BCS theories.  We are deliberately studying this idealized model as a paradigmatic example of a low-density system with strong spin-orbital effects where sharp conclusions can be made. 
First, we show that the original FE-QCP is preempted by a {\it ferroelectric density wave} (FDW) QCP due to the coupling between Dirac fermions and transverse optical phonons. This FDW state breaks translational symmetry in addition to inversion and rotational symmetries. Second, we find that the FDW fluctuations, which are peaked on a spherical surface in momentum space, mediate a much stronger attractive interaction as compared to the fluctuations of the uniform FE order. This attraction leads to a SC instability which, in turn, softens the FDW fluctuations  and enhances the pairing interaction even further. Consequently, because of this feedback effect, the system undergoes a first-order transition into a zero-density superconductor coexisting with ferroelectricity of some sort (finite-momentum or uniform), before the putative FDW-QCP is reached. Thus,
in addition to obtaining a zero-density pairing instability, we find an effective \textit{attraction} between FDW and SC order parameters, in sharp contrast to the usually observed competition between SC and other types of order such as antiferromagnetism, nematicity, or charge density wave~\cite{Balseiro1979,Das2006,Vorontsov2010,Moon2012,Fernandes2010}.

{\it Model.---}We start with the effective low-energy model for a polar Dirac semimetal near a FE transition derived in Ref.~\cite{KoziiBiRuhman2019}.  The direct Coulomb repulsion is strongly screened by the longitudinal optical phonon mode which remains massive at the critical point and does not contribute to the low-energy properties~\cite{AMbook}. The massless 3D  Dirac fermions  $\psi_\bk$ are then only coupled to the transverse optical phonon mode $\varphi_i(\bq) = P_{ij}(\bq)u_j(\bq)$, where $u_i(\bq)$ is proportional to the optical phonon displacement, $P^{ij}(\bq) = \delta^{ij} - \hat q^i \hat q^j$  is the projector onto the transverse modes, and $\hat \bq \equiv \bq/|\bq|$. The imaginary-time action of the system is given by $S = \int d\tau (\mathcal L_\psi + \mathcal L_\varphi + \mathcal L_{\psi-\varphi})$, with 
\begin{align} \label{Eq:S}
&\mathcal  L_{\psi} =   \sum_{n=1}^N \sum_{\bk}    \psi^\dagger_n(\tau,\bk) \left( \partial_\tau + iv_F \gamma^j \gamma^0  k_j\right)\psi_n(\tau, \bk), \nonumber \\ 
&\mathcal L_\varphi = \frac12 \sum_{\bq} \varphi^*_{i}(\tau,\bq) \left( -\partial_\tau^2 + c^2 q^2 +r\right) \varphi_{i}(\tau,\bq),  \\ 
&\mathcal  L_{\psi-\varphi} = \lambda  \sum_{n=1}^N\sum_{\bk \, \bq}   \psi^\dagger_n (\tau, \bk + \bq) \gamma_i \psi_n(\tau,\bk) \varphi_{i}(\tau, \bq) . \nonumber
\end{align}
The first term describes $N$ Dirac points and contains the Hermitian $\g$-matrices $\gamma_0 = \sigma_x\otimes s_0$ and $\gamma_i = \sigma_y \otimes s_i$ ($i=x,y,z$), where $\sigma_i$ ($s_i$) are Pauli matrices in orbital (spin) space, and $\sigma_0$ and $s_0$ are the identity matrix. While $N=4$ in cubic systems such as PbTe, we consider the formal limit $N\gg 1$ to neglect vertex corrections in our calculations. The second term describes the propagating transverse optical phonons, with $c$ denoting the phonon velocity and $r$ being the bare phonon mass. Note that $r = 0$ marks the bare FE-QCP~\cite{firstfoot}. Finally, the last term contains the coupling constant $\lambda$ between the Dirac electrons and the transverse optical phonons.

%Note that here we have assumed the mass term is tuned to zero uniformly. By that we have assumed the system is clean. How mass disorder effects our results is an important question, which is relevant to experiments.

{\it Tendency to FDW order.---}In the normal state, the coupling to the Dirac electrons leads to a renormalization of the phonon spectrum encoded in the polarization operator:
\be  
\Pi_n^{ij}(i \Omega, \bq) = \tl^2 \alpha^2 (v_F^2 q^2  - \Omega^2) \ln\frac{v_F\Lambda_0 e^{1/3}}{\sqrt{v_F^2 q^2 + \Omega^2}} P^{ij}(\bq), \label{Eq:Pibare}
\ee 
where $\tl^2 \equiv \lambda^2 N/12 \pi^2 v_F c^2$, $\alpha \equiv c/v_F$ is the velocity ratio, and $\Lambda_0$ is the high-momentum cutoff. We assume throughout this work that the effective coupling constant is small, $\tl \ll 1$, and that Fermi velocity is much larger than the bare transverse phonon velocity, $\alpha\ll1$. The effective normal state  phonon propagator is then given by $\hat D_n^{-1} = \hat D_0^{-1} - \hat  \Pi_n$, where $(D_0^{-1})_{ij}(i\Omega, \bq) = (r + c^2 q^2 + \Omega^2)P_{ij}(\bq)$ is the bare  propagator~\cite{secondfoot}. Minimization with respect to momentum reveals that, due to the logarithmic dependence in Eq.~\eqref{Eq:Pibare}, the original $q=0$ FE transition is preempted by one at a finite momentum $|\bs q| = Q$, where
\be
Q = \Lambda_0 \exp\left( - \frac1{\tl^2} - \frac16 \right), \label{Eq:Q0}
\ee
and the static polarization is transverse.
Near the minimum $q=Q$, the static bosonic propagator can be expanded as
\be
D_n^{-1}(0,q) \approx  r - r_{\text{FDW}} + \tl^2 c^2 (q
-Q)^2, \label{Eq:Vapprox}
\ee
where $r_{\text{FDW}} = \tl^2 c^2 Q^2/2$. We note that $Q$ is exactly the scale appearing in the RG equations derived in Ref.~\cite{KoziiBiRuhman2019} where the system reaches the strong-coupling regime. Equation~\eqref{Eq:Vapprox} has a minimum on the whole sphere $q=Q$ instead of a single point at $q=0$ implying more phase space for fluctuations. Bosonic models of this type are known to undergo a weak fluctuation-driven first-order transition~\cite{Brazovskii1975,Fernandes2008}, which we will neglect hereafter.

%\footnote{To properly invert the bosonic propagator, one should keep the mass of the longitudinal mode large but finite and set it to infinity only at the end of the calculation.}

{\it Synergistic FDW and SC orders.---}We proceed to discuss the possibility of a pairing instability  near the emergent FDW-QCP by solving the coupled Eliashberg equations~\cite{SM} inside the paraelectric (PE) phase at the temperature $T = 0$. First, we analytically solve the equations in a BCS-like approximation which implies neglecting the electronic normal self-energy as well as the frequency and momentum dependence of the superconducting order parameter $\Delta$. The boson self-energy $\Pi(i\Omega,q,\Delta)$, however, is computed fully self-consistently. Then, we compare our analytic result with a numerical solution of the frequency-dependent Eliashberg equations and find good qualitative agreement. As we shall shortly detail, we find a range of the parameter $r$, where the gap equation has a nontrivial solution $\Delta = \Delta_{*}(r) \ne 0$ in spite of the vanishing density of states. However, by analyzing the free energy in the vicinity of this solution,  we show that $\Delta_{*}$ is in fact an unstable solution, which indicates the existence of a stable minimum where the system develops both FDW and SC orders simultaneously through a first-order phase transition.

\begin{figure}
  \centering
  \includegraphics[width=1.\columnwidth]{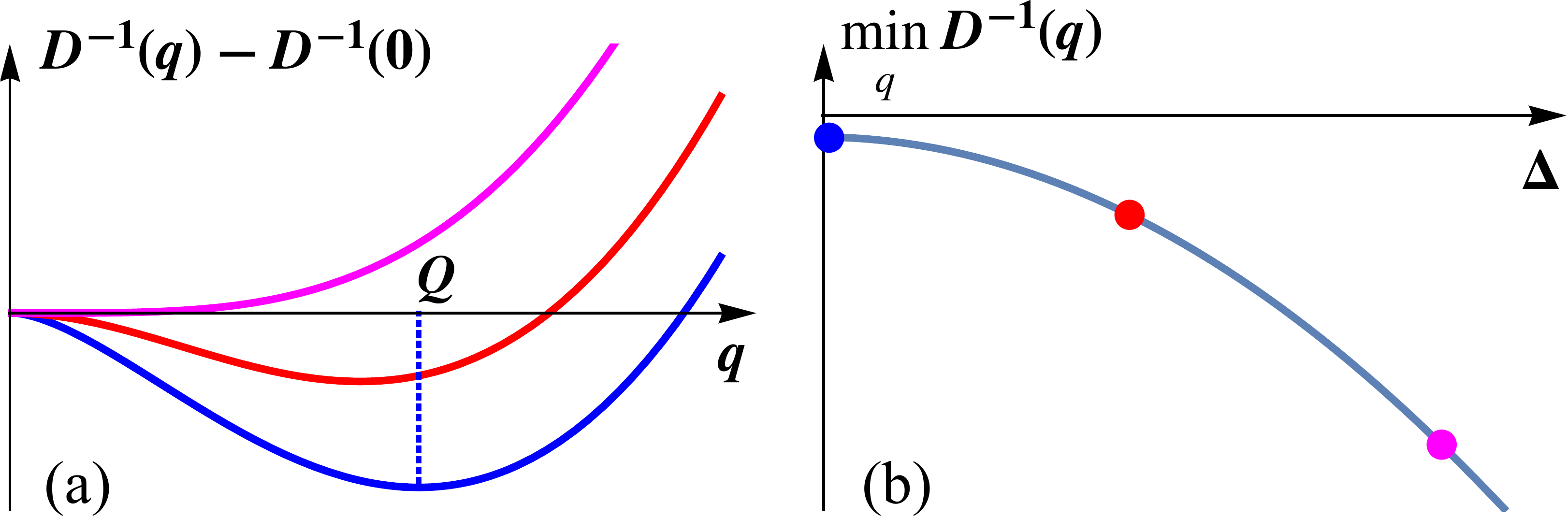}
    \caption{ (a) Inverse static phonon propagator $D^{-1}(q)$, which is proportional to the inverse pairing interaction, as a function of momentum for fixed $r$ and $\Delta/\Delta_0 = 0$ (blue), 0.5 (red), and 1 (purple), with $\Delta_0$ given by Eq.~\eqref{Eq:Delta0}. $D^{-1}(q=0)$ was subtracted for better visualization. (b) The minimal value of $D^{-1}$ shifts to lower values with increasing $\Delta$, signaling a decrease in the spectral mass $m^2(r,\Delta)$ from Eq.~\eqref{eq:ms}.}
  \label{Fig:Veff}
\end{figure}

The mechanism driving the first-order transition stems from the feedback of the
SC gap $\Delta$ on the  boson propagator $D(i\Omega, q, \Delta)$. For $\Delta=0$, the one-loop polarization operator is given by Eq.~\eqref{Eq:Pibare} and the minimum of $D^{-1}$ is at the finite momentum $q=Q$, Eq.~\eqref{Eq:Q0}. The introduction of a finite (constant) $\Delta$ has two important effects on $D^{-1}(i\W,\bq,\D)$~\cite{SM}, which  we illustrate in Fig.~\ref{Fig:Veff}. First, the position of the minimum gradually shifts toward smaller $q$ until it  merges with the $q=0$ local maximum at $\Delta_0$ given by
\be  
\Delta_0 = \exp(-5/6) v_F Q. \label{Eq:Delta0}
\ee 
When $\Delta > \Delta_0$, the minimum of $D^{-1}$ shifts back to $q=0$. Second, a finite $\Delta$ reduces the value of the  global minimum of $D^{-1}$, as shown in Fig.~\ref{Fig:Veff}(b). Thus, increasing $\D$ makes the FDW mode softer, which in turn increases the pairing interaction. This scenario should be contrasted with other models for superconductivity in the vicinity of a QCP, where the SC order and the order associated with the QCP generally compete, see, e.g., Refs.~\cite{Balseiro1979,Das2006,Vorontsov2010,Moon2012}. Here, the two long-range ordered states not only mutually enhance each other, but they are also characterized by the same energy scale $\vf Q$.

\begin{figure}
  \centering
  \includegraphics[width=0.8\columnwidth]{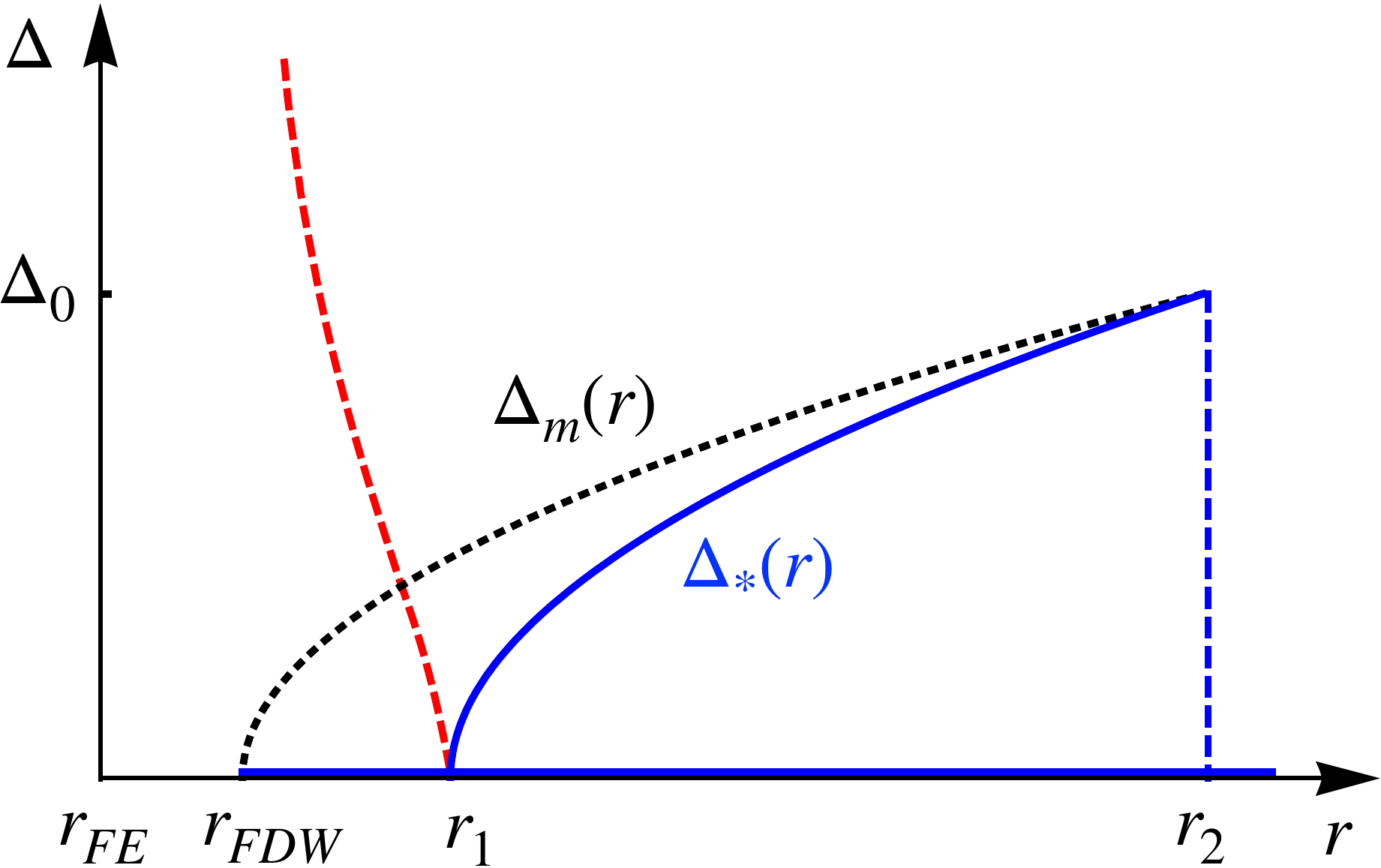}
    \caption{Nontrivial superconducting solution $\D_*$ (blue line) as a function of the tuning parameter $r$. The red dashed line represents the ``superconducting dome'' one would obtain by ignoring the feedback effect of the SC gap $\Delta$ on the phonon propagator $D$. The values of $\Delta_0$, $r_{\text{FDW}}$, $r_1$, and $r_2$ are defined in the main text. The black dotted line shows $\Delta_m(r)$ defined below Eq.~\eqref{eq:ms}. The exponentially small region $r_{\text{FDW}} < r<r_1$ was inflated for illustrative purposes.}
  \label{Fig:Delta}
\end{figure}

\footnotetext[6]{By projecting the interaction in Eq.~\eqref{Eq:S} onto the  momentum-independent pairing channels we find that the only attractive channel is the fully symmetric $s$-wave~\cite{SM}.}

{\it Gap Equation.---}We now  discuss the solutions of the gap equation when approaching the FDW-QCP from the PE side, focusing on the leading $s$-wave channel~\cite{Brydonetal2014,Note6}:
\be
\Delta= \frac{6 \tl^2 \alpha^2 v_F^3}{\pi N} \int_{-\infty}^{\infty} d\omega \int_0^{\Lambda_0} dk k^2  \frac{\Delta \cdot D(i \omega ,k, \Delta)}{\omega^2 + v_F^2 k^2 + \Delta^2}. \label{Eq:gapsimple}
\ee
Note that $D$ plays the role of an effective pairing interaction, and that this equation always has the trivial solution $\Delta=0$.  Furthermore, this equation admits a nontrivial solution because the integral gives a Cooper-like logarithm. The one-dimensional nature of this logarithm, however, does not arise from the electron's propagator having a finite chemical potential, but instead from the phonon's propagator being peaked at a finite momentum~\cite{SM}. If we neglect the feedback of $\Delta$ on $D$ discussed above, we find a ``standard'' instability toward pairing peaked at the FDW-QCP. As shown by the red dashed line in Fig.~\ref{Fig:Delta}, in this case $\Delta$ diverges at the QCP $r=r_{\mathrm{FDW}}$. However, upon including $\Delta$ self-consistently in $D$, the situation changes dramatically. Indeed,  solving Eq.~\eqref{Eq:gapsimple} reveals three important ranges of $r$. In the immediate vicinity of the putative FDW-QCP, $r_{\text{FDW}} < r \le r_1$, the only solution we find is $\Delta=0$.  The first nontrivial solution $\Delta_*$ appears at $r_1 = r_{\text{FDW}} + \delta r$, where
\begin{align} 
\delta r \sim \tl^2 c^2 Q^2 \, &\exp\left( -\frac{N\sqrt{1+\alpha^2}}{6 \tl \alpha} + \frac2{\tl} \right). \label{Eq:deltar}
\end{align}
The value of $\Delta_*$ increases as we move further away from the QCP, $r_1 < r < r_2$, until it reaches the value $\Delta_* = \Delta_0$ given by Eq.~\eqref{Eq:Delta0} at  $r=r_2$:
\be  
r_2 = 6e^{-5/3}\left( 1+\frac{\tl^2}2 \right) c^2 Q^2. \label{Eq:r2}
\ee 
Finally, the nontrivial solution disappears abruptly for $r \ge r_2$ 
provided that $12\alpha/N \lesssim 1$. The full dependence of the  solution $\Delta_*(r)$ on the tuning parameter $r$ is shown in Fig.~\ref{Fig:Delta} by the solid blue curve. Note however, that the region $r_{\text{FDW}} < r < r_1$ is exponentially small even on the scale of the exponentially small $\vf Q$.

To confirm that these results are not an artifact of the BCS approximation, we also numerically solve the coupled Eliashberg equations with full frequency dependence of the gap function and including the normal part of the electron's self-energy~\cite{SM}. 
Overall, we find the same qualitative behavior of a nontrivial solution that grows upon increasing $r$ away from $r_{\mathrm{FDW}}$. The main impact of the normal part of the self-energy is a reduction in the value of $\Delta_*$.

\begin{figure}
  \centering
  \includegraphics[width=0.8\columnwidth]{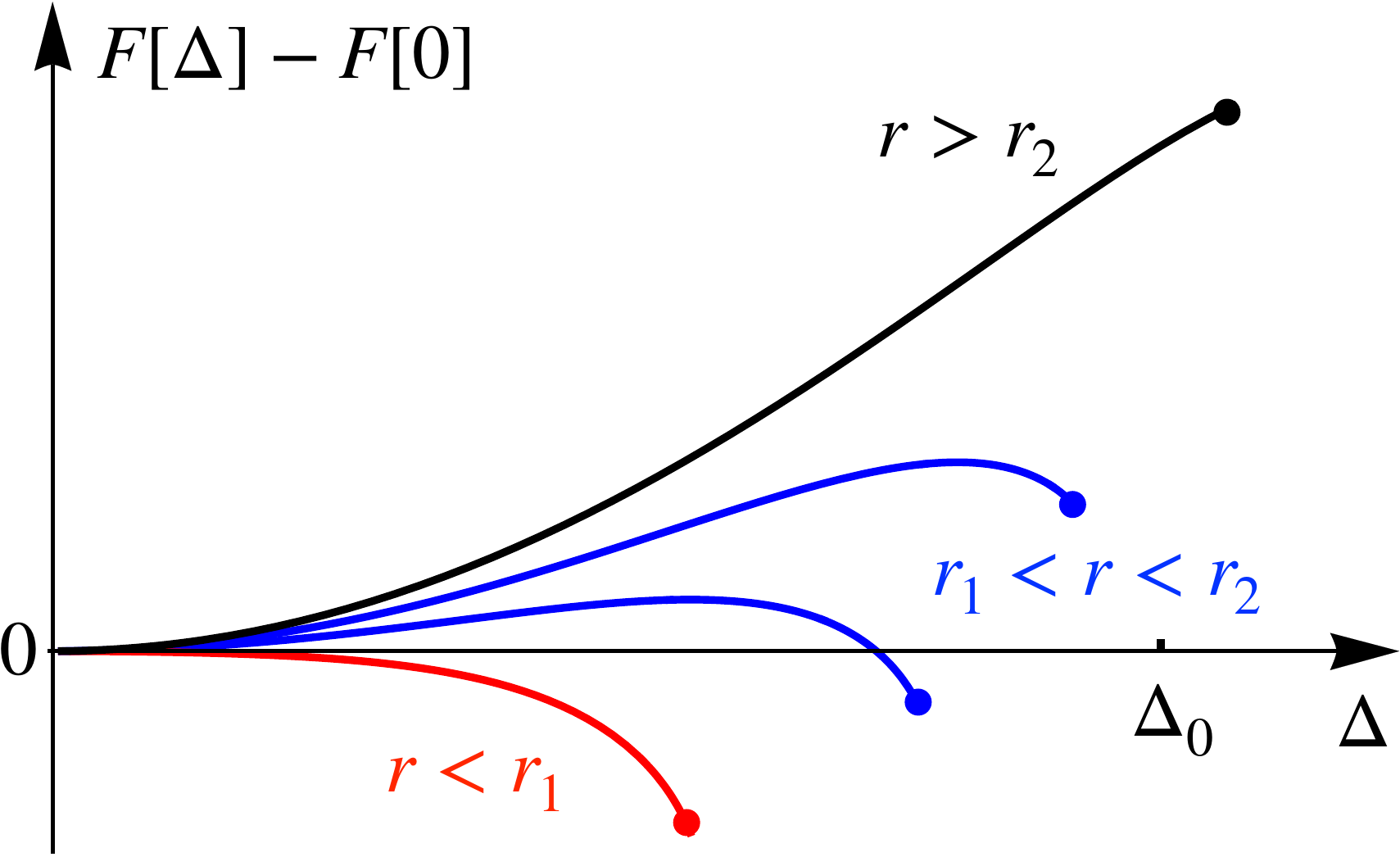}
    \caption{ Superconducting free energy $F[\Delta]$ for different fixed values of the tuning parameter $r$ in the PE phase. The red curve with a maximum at $\Delta=0$ is from the region $r_{\text{FDW}} < r < r_1$. The blue curves are from the region $r_1 < r < r_2$ and exhibit a local minimum at $\Delta=0$ and a local maximum at $0 < \Delta_* < \Delta_0$. The black curve with $r > r_2$ has a single minimum at $\Delta=0$. All curves end at the points where the corresponding spectral mass $m^2(r,\Delta)$ vanishes (bold dots), i.e., at $\Delta_m(r)$, which indicates an instability toward the FDW state.} 
  \label{Fig:Freeenergy}
\end{figure}

{\it First-order transition into an FDW$+$SC state.---}The increase of the nontrivial solution $\Delta_*(r)$ upon moving away from the QCP in the range $r_1<r<r_2$ seems at odds with the na\"{i}ve expectation that superconductivity is maximal at the QCP, or at least grows upon approaching it. To understand this, we analyze the superconducting free energy $F[\Delta]$.  Expressing the gap equation~\eqref{Eq:gapsimple} as $1 = f[\Delta_*]$, the free energy derivative can be conveniently written as $\partial F[\Delta]/\partial \Delta = \Delta - \Delta f[\Delta]$. Because $\partial f / \partial \Delta > 0$ at $\D=\D_*$ (as can be understood from the dependence of $D$ on $\Delta$ shown in Fig.~\ref{Fig:Veff}), it follows that $\partial^2 F/\partial \D^2<0$ at $\D=\D_*$, i.e., the nontrivial solution is {\it unstable}~\cite{SM}. The fact that $\Delta_*$ is a local maximum of the free energy is further illustrated in Fig.~\ref{Fig:Freeenergy}, where $F[\Delta]$ is plotted as a function of $\Delta$ for different fixed values of $r$. This situation should be contrasted with the standard BCS gap equation for a Fermi liquid with attractive interactions, where the nontrivial solution is always a local minimum of the free energy.

To understand the relationship between the SC and FDW order parameters, we  introduce the phonon spectral mass:
\be 
m^2(r,\Delta) \equiv {\min_q D^{-1}(0,q,\Delta\,, r)}. \label{eq:ms}
\ee
The PE phase corresponds to $m^2(r,\Delta) >0$, while $m^2(r,\Delta) <0$ indicates the structural instability toward FDW order. 
The equation $m^2 (r,  \Delta_m)=0$ defines the threshold gap value $\Delta_m(r)$ above which the PE-superconducting system is unstable toward the development of FDW order in conjunction with superconductivity for a given $r$. These values correspond to the termination points of the free energy curves in Fig.~\ref{Fig:Freeenergy}. The curve $\Delta_m(r)$ is shown by the black dotted line in Fig.~\ref{Fig:Delta}, with  $\D_m(r_{\mathrm{FDW}})  = 0$ and $\D_m(r_2) = \D_0$.

We thus conclude that the solution $\D_*(r)$ is an instability line  separating the two stable solutions: (i) A trivial $\D = 0$ solution and (ii) a second nontrivial solution $\D > \D_m(r)$ inside the FDW phase, which cannot be accessed by our gap equation derived in the PE phase. For $r \le r_1$, the trivial solution $\D=0$ is unstable and superconductivity becomes inevitable. We note that we cannot determine whether the ordered state has uniform FE or modulated FDW order, since our analysis is done on the PE side of the transition. Nonetheless, we refer to the ordered state as FDW for definiteness.

The shape of the free energy landscape in Fig.~\ref{Fig:Freeenergy} is typical of a system undergoing a first-order transition, as it displays two local minima (one at $\D = 0$ and another at $\D > \D_m(r)$) and one local maximum at $0 < \D_*(r) < \D_m(r)$ for $r_1 < r < r_2$. This behavior is rooted in the unusual positive feedback effect of a gap opening on the effective pairing interaction $D$ of Eq.~\eqref{Eq:gapsimple}~\cite{thirdfoot}. This feedback is  stabilized only when FDW order sets in, signaling a coexistence between SC and FDW orders. This also explains the absence of a nontrivial solution $\D_*(r)$ for $r>r_2$, since in this range the line $\D_*(r)$ is above $\D_m (r)$, which means that the system must already be in the FDW phase.

%\footnote{Increasing $\Delta$ generally has two opposite effects on the boson propagator: (1) It shifts the minimum of $D^{-1}$ toward smaller momenta $q$  and (2) It decreases the phonon spectral mass. The first effect reduces the strength of the FDW fluctuations while the second one increases it, and our calculation shows that the latter effect is dominant.}

\begin{figure}
  \centering
  \includegraphics[width=0.8\columnwidth]{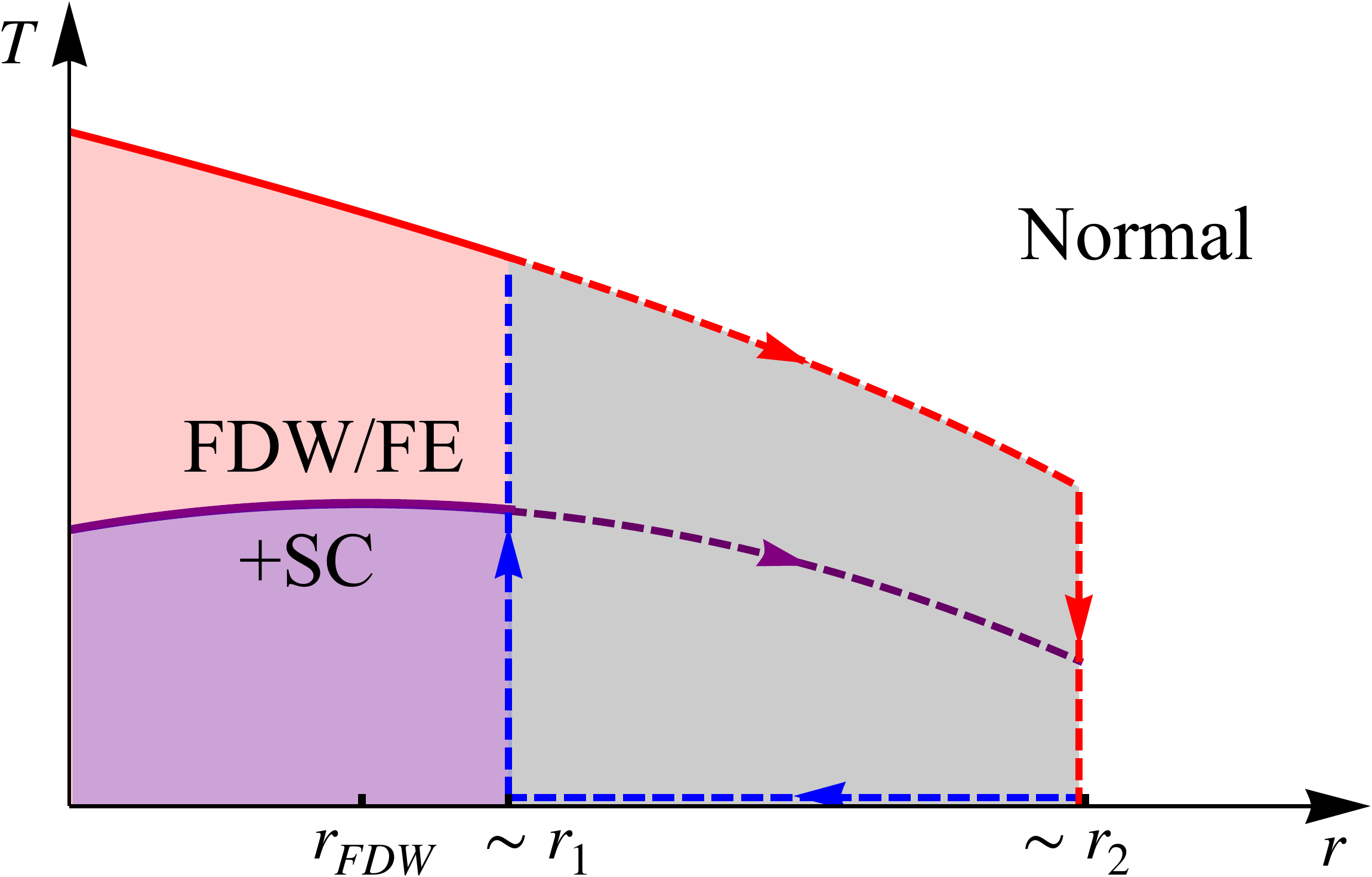}
    \caption{ The expected phase diagram showing the interplay between the FDW/FE and SC orders. The two ordered states enhance each other upon approaching the bare critical point $r_{\text{FDW}}$, resulting in a  first-order transition into the state where FDW/FE and SC coexist. The transition is accompanied by the characteristic  bistable region with two local energy minima (gray region). ``Normal'' marks  the nonsuperconducting  paraelectric region.}  \label{Fig:pd}
\end{figure}

 The expected first-order phase diagram emerging from this analysis is shown schematically  in Fig.~\ref{Fig:pd}. Upon approaching the FE-QCP from large values of $r$, the system can remain in the non-SC and PE state, which is locally stable. At some point $r\ge r_1$, $\D$ jumps to a finite value and the system undergoes a first-order transition into the state where SC and FDW coexist. 
On the other hand, upon increasing $r$ from the FDW side ($r<r_{\mathrm{FDW}}$), the system is already SC, and therefore,  can remain in this locally stable minimum until some larger value of $r$ is reached, where $\D$  jumps abruptly to zero and the system becomes PE. A detailed calculation of the Ginzburg-Landau free energy in the ordered FDW state, which is needed to determine precisely the global free-energy minimum, is outside the scope of this work. Note also that our approach is justified as long as the first-order simultaneous SC+FDW transition happens before the fluctuation-driven weakly first-order transition expected for the purely bosonic FDW propagator of Eq.~\eqref{Eq:Vapprox}.

In order for the superconducting state discussed above to be stable against phase fluctuations, it must have a finite superfluid stiffness $\rho_s$.
In a BCS superconductor, $\rho_s$ is proportional to the electronic density. In the case of a Dirac semimetal at charge neutrality, we find that the gap itself generates an emergent density scale $(\D/v_F)^3$, which leads to a finite  stiffness despite the vanishing density of states. In the uniform FE phase, we find 
\be
\rho_s \sim \frac{e^2}{v_F} \Delta^2 \ln\left(\frac{v_F \Lambda_0}{\Delta}\right)\,.
\ee
We expect this result to hold inside the FDW phase as well.  The finite stiffness even at charge neutrality originates from the charge reservoir provided by the filled valence band~\cite{ChubukovEreminEfremov2016}.

{\it Discussion.---}
We have found that the strong coupling between a FE QCP and a charge neutral Dirac point~\cite{KoziiBiRuhman2019} leads to the  first-order transition into the state where  nonuniform ferroelectric order and superconductivity appear simultaneously. The key ingredients crucial for this result are the Dirac band touching and its linear coupling to the inversion-odd transverse critical modes.

Our low-energy model of a Dirac semimetal near a putative QCP is possibly  relevant to the alloy Pb$_{x}$Sn$_{1-x}$Te. SnTe ($x=0$) is a ferroelectric crystalline topological insulator~\cite{hsieh2012topological}. PbTe ($x=1$), on the other hand, is a paraelectric and higher order topological insulator~\cite{Robredo2019}. For intermediate $x$, a ferroelectric-paraelectric transition takes place at $x\approx0.5$~\cite{Daughton1978,ElectronicandDynamicalBook}. Doping this composition with indium promotes superconductivity, and is accompanied by unusual features in the phonon density of states ~\cite{Parfenev2001,Sapkota2020,ran2018unusual}.  Using experimental parameters for PbTe we estimate  the effective coupling constant to be $\tl^2 \approx 0.27$~\cite{SM}.  Thus, we predict the possibility of nonuniform ferroelectricity and low-density superconductivity in the alloy $(\mathrm{Pb}_{1-x}\mathrm{Sn}_{x})_{1-y}\mathrm{In}_y$Te. 

The onset of an FDW phase in this compound should lead to translational symmetry breaking, which could be experimentally detected in the lattice degrees of freedom by neutron and Raman spectroscopy, and in the electronic degrees of freedom by angle-resolved photoemission spectroscopy and quantum oscillations~\cite{LiangOng2017}.

Furthermore, our results raise  a number of questions. The first challenge is to perform a complimentary study on the ordered side of the transition. Second, the phenomenological nature of the zero density superconductor calls for research, e.g., the nature of its collective modes. Also, studying the influence of disorder on the coupled transition is critical in making connection to real materials. Finally, understanding how the phase diagram depends on chemical potential may help make connection to ferroelectric metals with a Fermi surface.  For instance, in the case of a Fermi liquid near a putative dipolar ferromagnetic QCP (rather than our dipolar ferroelectric QCP), it is well established that an instability toward a finite-momentum magnetic state is driven by quantum fluctuations~\cite{Belitz1997,Maslov2009,Green2018}. 
%Moreover, in some of these systems, the electronic coupling to two phonon processes~\cite{Ngai,kumarquasiparticle} has been discussed~\cite{vandermarelpossible,volkov2021superconductivity,kiseliov2021theory}. While this coupling gives a good estimates for $T_c$ at moderate to strong coupling strength~\cite{vandermarelpossible,volkov2021superconductivity,kiseliov2021theory}, it is irrelevant in the RG sense~\cite{volkov2021superconductivity}. It therefore poses an interesting question of whether such a coupling can modify the critical properties of the FE-QCP discussed here.  

\begin{acknowledgements}
We acknowledge helpful discussions with A.~Chubukov, M.~Gastiasoro, D.~Maslov, and T.~Trevisan. V.K. was supported by  the Quantum Materials program at LBNL, funded by the U.S. Department of Energy under Contract No. DE-AC02-05CH11231. Work by A.K. (while at the University of Minnesota) and R.M.F. was supported by the U.S. Department of Energy through the University of Minnesota Center for Quantum Materials, under Grant No.~DE-SC-0016371. The final stage of the work by A.K. and J.R. was supported by the Israel Science Foundation (ISF), and the Directorate for Defense Research and Development (DDR\&D), Grant No. 3467/21. The final stage of the work by V.K. was performed in part at Aspen Center for Physics, which is supported by National Science Foundation Grant No. PHY-1607611 and by a grant from the Simons Foundation. 
\end{acknowledgements}

\bibliography{zerodensity}

\begin{thebibliography}{75}
\bibitem [{fir()}]{firstfoot}%
  \BibitemOpen
  \href@noop {} {\bibinfo  {journal} {Note that here we have assumed the mass
  term is tuned to zero uniformly. By that we have assumed the system is clean.
  How mass disorder effects our results is an important question, which is
  relevant to experiments}\ }\BibitemShut {NoStop}%
\bibitem [{sec()}]{secondfoot}%
  \BibitemOpen
\bibfield  {journal} {  }\href@noop {} {\bibinfo  {journal} {To properly invert
  the bosonic propagator, one should keep the mass of the longitudinal mode
  large but finite and set it to infinity only at the end of the calculation}\
  }\BibitemShut {NoStop}%
\bibitem [{thi()}]{thirdfoot}%
  \BibitemOpen
  \href@noop {} {\bibinfo  {journal} {Increasing $\Delta$ generally has two
  opposite effects on the boson propagator: (1) It shifts the minimum of
  $D^{-1}$ toward smaller momenta $q$ and (2) It decreases the phonon spectral
  mass. The first effect reduces the strength of the FDW fluctuations while the
  second one increases it, and our calculation shows that the latter effect is
  dominant}\ }\BibitemShut {NoStop}%
\end{thebibliography}

\newpage

\begin{widetext}

\begin{center}
\textbf{\large Supplemental Materials for ``Synergetic ferroelectricity and superconductivity in zero-density Dirac semimetals near quantum criticality''} 
\end{center}
%%%%%%%%%% Merge with supplemental materials %%%%%%%%%%
%%%%%%%%%% Prefix a "S" to all equations, figures, tables and reset the counter %%%%%%%%%%
\setcounter{equation}{0}
\setcounter{figure}{0}
\setcounter{table}{0}

\makeatletter
\renewcommand{\theequation}{S\arabic{equation}}
\renewcommand{\thefigure}{S\arabic{figure}}
\renewcommand{\thetable}{S\Roman{table}}

In this Supplemental Materials we discuss the technical details of the analytical and numerical solutions for the system of Eliashberg equations and discuss their stability. In Sec.~I, we present a model to study and write down its effective low-energy action in Nambu space. In Sec.~II, we analyze possible symmetries of the gap functions. In Sec.~III,  we write down a general system of coupled Eliashberg equations. We solve this system analytically within the BCS approximation in Sec.~IV and numerically in Sec.~V taking into account full frequency dependence. In Sec.~VI, we speculate on possible scenarios for the free energy flow in the ferroelectric (FE) phase. Finally, in Sec.~VII, we estimate the value of the effective coupling constant in PbTe/SnTe.

\section{I. Model and effective action}

We consider a model for a massless Dirac fermion in a polar crystal with a chemical potential exactly at the charge neutrality point coupled to the nearly critical transverse optical phonon mode. The soft phonon modes are associated with the transition into the ferroelectric phase. The low-energy effective imaginary-time action at zero temperature, when rewritten in Nambu space, takes the form~\cite{KoziiBiRuhman2019}

\be
S= S_{\psi} + S_{u} + S_{\psi-u},
\ee
with

\begin{align}
S_{\psi} &=  \frac12 \sum_{n=1}^N \int \frac{d \omega \, d^3 \bk}{(2 \pi)^4}   \Psi^\dagger_n(\omega, \bk)\left[-i \omega \tau_0 + i v_F \gamma^j \gamma^0 \tau_z  k_j \right]\Psi_n(\omega, \bk), \nonumber \\ S_u &= \frac{1}{2} \int \frac{d \Omega \, d^3 \bq}{(2 \pi)^4} (\Omega^2 + c^2 q^2 + r) u_i(\Omega, \bq) u_j^*(\Omega, \bq) P^{ij}(\bq), \nonumber \\ S_{\psi-u} &= \frac{\lambda}2 \sum_{n=1}^N \int \frac{d \Omega \, d \omega \, d^3 \bk \, d^3 \bq}{(2 \pi)^8} \Psi^\dagger_n (\omega + \Omega, \bk + \bq) \gamma_i \tau_z \Psi_n(\omega, \bk) u_j(\Omega, \bq) P^{ij}(\bq), \label{SMEq:S}
\end{align}
where the summation over the repeated spatial indices $i, j = x,y,z$ is implied. Here we introduced the Nambu spinor according to

\be
\Psi(\omega, \bk) = \left( \begin{array}{c}  \psi_{\uparrow}(\omega, \bk) \\ \psi_{\downarrow}(\omega, \bk) \\ \psi^\dagger_{\downarrow}(-\omega, -\bk) \\ -\psi^\dagger_{\uparrow}(-\omega, -\bk)   \end{array}  \right)_N. \label{SMEq:spinordef}
\ee
In this action, $\tau_i$ are the Pauli matrices in Nambu space ($\tau_0$ is the identity matrix), $n$ numerates different electron flavors, $u_i(\omega, \bq)$ is the lattice displacement field which corresponds to the optical phonon modes, and $P^{ij}(\bq) = \delta^{ij} - \hat q^i \hat q^j$  is the projector onto the transverse direction, where $\hat\bq = \bq / q$. Parameter $r$ is the bare phonon mass that controls the proximity to the bare critical point and the $\gamma$-matrices are chosen to be

\be
\gamma_0 = \sigma_x \otimes s_0, \qquad \gamma_x = \sigma_y \otimes s_x, \qquad \gamma_y = \sigma_y \otimes s_y, \qquad \gamma_z = \sigma_y \otimes s_z, \label{SMEq:gammadef}
\ee
where $\sigma_i$ and $s_i$ are the Pauli matrices in the orbital and spin space, respectively, while $\sigma_0$  and $s_0$ are the identity matrices. The inversion operator in this basis is given by $\mathcal P = \gamma_0$.  The additional structure of the spinor~\eqref{SMEq:spinordef} in the orbital space $\sigma$ and ``flavor'' space $n$ is implicit.

Following the result of Ref.~\cite{KoziiBiRuhman2019}, we assume that the direct Coulomb repulsion between electrons is screened by the longitudinal optical phonons which remain massive at the transition because of the splitting between the longitudinal and transverse optical modes. Consequently, both Coulomb repulsion and longitudinal phonon mode are absent in the effective low-energy action.

\section{II. Gap functions}
Here we discuss the symmetries of the superconducting (SC) order parameter. In particular, we show that the leading instability is the simplest (fully symmetric) $s$-wave state which we use in our BCS and Eliashberg analyses later.  

The superconducting order parameters with zero total momentum generically have form
\be
\hat \Delta_M (\bk) = \psi_{-\bk,\alpha}\,\mathcal M_M^{\alpha \beta}(\bk)\psi_{\bk,\beta},
\ee
where $\mathcal M_M(\bk)$ is a matrix encoding the spin-orbital structure of the pairing state $M$ as well as its momentum dependence. The antisymmetric property of the pair wave function implies that $\mathcal M^T_M(-\bk) = - \mathcal M_M(\bk)$. We are interested in the states that open a spectral gap at charge neutrality, consequently, below we focus on the momentum-independent pairing channels:

\begin{align}
    &\mathcal M_{I} = -i s_y\\
    &\mathcal M_0 = -i s_y\gamma_0 \nn\\
    &\mathcal M_5 = -i s_y\gamma_5 \nn \\
    &\mathcal M_j = -i s_y\gamma_j\;;\;j=x,y,z \nn
\end{align}
where $\gamma_5 = \gamma_0 \gamma_x \gamma_y \gamma_z = -\sigma_z \otimes s_0$.

To identify the leading instability, we decompose  the interaction into these pairing channels. First, we integrate out the phonons and write down the resulting interaction as
\be
H_{\text{int}} = -\frac{\lambda^2}2\sum_{\bp \bk \bq} D(q)P_{ij}(\bq)\gamma_i^{\alpha \delta}\gamma_j^{\beta \gamma} \psi^\dag _{\bk + \bq,\alpha}\psi_{\bk,\delta}\psi^\dag_{\bp - \bq,\beta}\psi_{\bp,\gamma}\,,
\ee
where $\lambda^2 D(q)$ is the effective attractive interaction potential, $D(q)$ is the phonon propagator, and the summation over the repeated spatial ($i,j=x,y,z$) and spinor ($\alpha,\beta,\gamma,\delta$) indices is implied. Focusing on the Cooper channels with zero total momentum, the interaction term can be rewritten as a pairing Hamiltonian:

\be
H^{\text{p}}_{\text{int}} = -\frac{\lambda^2}4\sum_{\bp \bk} \left[ D(\bp - \bk)P_{ij}(\bp - \bk)\gamma_i^{\alpha \delta}\gamma_j^{\beta \gamma} - D(\bp + \bk)P_{ij}(\bp + \bk)\gamma_i^{\alpha \gamma}\gamma_j^{\beta \delta}\right] \psi^\dag _{\bp,\alpha}\psi^\dag_{-\bp,\beta}\psi_{-\bk,\gamma}\psi_{\bk,\delta}\,,
\ee
where we have antisymmetrized the whole expression with respect to indices $\gamma$ and $\delta$ for convenience. Next, we assume that the effective interaction is dominated by its $s$-wave harmonic, i.e., $D(\bp \pm \bk) \approx D_s(p,k)$. Finally, to decompose the interaction into the momentum-independent pairing channels, we use the identity 

\be
\int \frac{d\W_{\bp} d\W_{\bk}}{(4\pi)^2}\left[ P_{ij}(\bp-\bk)\gamma_i^{\alpha \delta}\gamma_j^{\beta \gamma} - P_{ij}(\bp+\bk)\gamma_i^{\alpha \gamma}\gamma_j^{\beta \delta} \right]=  \mathcal M_{I}^{\dag \alpha \beta} \mathcal M_{I}^{\gamma \delta} - \mathcal M_0^{\dag \alpha \beta} \mathcal M_0^{\gamma \delta}-\mathcal M_5^{\dag \alpha \beta}\mathcal M_5^{\gamma \delta}  -\frac13 \mathcal M_j^{\dag \alpha \beta} \mathcal M_j^{\gamma \delta},
\ee
where $d\W_\bp = d(\cos \theta_\bp)d\phi_\bp$ and the summation over the repeated spatial indices $i,j=x,y,z$ is implied.  This identity is satisfied for any $p$ and $k$ and any spinor indices $\alpha,\beta,\gamma,\delta$. Consequently, the pairing interaction can be rewritten as 

\be 
H^{\text{p}}_{\text{int}} = -\frac{\lambda^2}4\sum_{\bp \bk} D_s(p,k)\sum_M a_M \hat \Delta_M^\dag(\bp) \hat \Delta_M(\bk) + \ldots,
\ee 
where the decomposition coefficients $a_M$ are given by 

\be  
a_I = 1, \qquad a_0 =a_5 = -1, \qquad a_j = -\frac13, \qquad j=x,y,z, 
\ee 
and $\ldots$ stands for the channels with higher angular momenta which we have neglected. 
Therefore, we find that the only attractive pairing channel is in the trivial $s$-wave representation $\mathcal M_{I}$, while all other momentum-independent channels are repulsive. Consequently, we focus on this channel hereafter. 

We also emphasize that the above consideration of the momentum-independent channels is mostly relevant for the charge neutrality point at $k=0$. At a finite doping when the Fermi surface is developed, some channels may merge into a single one after projecting onto the Fermi surface. For instance, $\mathcal M_{I}$ and $\mathcal M_{0}$ merge into a single $s$-wave channel, while $\mathcal M_{j}$ get admixtures of some momentum-dependent channels leading to different numbers $a_M$~\cite{KoziiFu2015,KoziiBiRuhman2019}. We still expect, however, that the leading pairing  instability remains in the $s$-wave channel~\cite{Brydonetal2014}.

\section{III. Eliashberg equations}

We proceed with writing down the system of coupled self-consistent Eliashberg equations for the electron and phonon Green's functions $\hat G(i\omega, \bk)$ and $\hat D(i\Omega, \bq)$, respectively. These are essentially the Dyson equations that neglect the vertex corrections. In our case, this approach is justified by the large number of electron flavors $N\gg1$ and the smallness of the velocities ratio $\alpha = c/v_F \ll 1$.

The Dyson equations for the propagators have the conventional form:

\be
\hat G^{-1} = \hat G_0^{-1} - \hat \Sigma, \qquad \hat D^{-1} = \hat D_0^{-1} - \hat \Pi,
\ee
where $\hat \Sigma$ and $\hat \Pi$ are the electron and phonon self-energies, correspondingly. The bare Green's functions are given by

\be
\hat G_0^{-1}(i\omega_n, \bk) = i \omega_n \tau_0 - i v_F \gamma^j \gamma^0 k_j \tau_z, \qquad D_0^{ij}(i\Omega_m, \bq) = \frac{P^{ij}(\bq)}{\Omega_m^2 + c^2 q^2 + r}, \label{SMEq:bareGreenfunctions}
\ee
where Matsubara frequencies are $\omega_n = \pi T (2n+1)$ and $\Omega_m = 2 \pi T m$ with integer $n$ and $m$. We note that the matrix structure of $\hat D$, $\hat D_0,$ and $ \hat \Pi$ is given by the projector $P^{ij}(\bq) = \delta^{ij} - \hat q^i \hat q^j$.

We parameterize the electron self-energy in Nambu space in the usual way:

\be
\hat \Sigma(i\omega_n, \bk) = i\omega_n \left[ I - \hat Z (i\omega_n, \bk) \right]\otimes \tau_0 + \hat\chi (i\omega_n, \bk)\otimes \tau_z + \hat \phi_1(i\omega_n, \bk) \otimes \tau_x + \hat \phi_2(i\omega_n, \bk) \otimes \tau_y,
\ee
where $I= \sigma_0 \otimes s_0$ is the identity matrix. Next, we do a number of approximations that allow us to treat the problem in a comprehensible way. First, according to the result from the previous section, we assume that the superconducting order parameter is real, has $s$-wave symmetry, and depends only on frequency but not on momentum, $\hat \phi_1(i\omega_n, \bk) = \phi(i\omega_n)\otimes I$, $\hat \phi_2(i\omega_n, \bk) = 0$. Second, we assume that the renormalization of the quasiparticle weight also depends on frequency only and is given by $\hat Z(i\omega_n, \bk) = Z(i\omega_n) \otimes I$. Finally, we assume that $\hat\chi (i\omega_n, \bk)$ only renormalizes the Fermi velocity. Hence, we use $v_F$ for the renormalized Fermi velocity and do not consider $\hat\chi (i\omega_n, \bk)$ hereafter. With these assumptions, the electron self-energy takes the form

\be
\hat \Sigma(i\omega_n) = i\omega_n \left[ 1 -  Z (i\omega_n) \right]\otimes \tau_0  +  \phi(i\omega_n) \otimes \tau_x.
\ee

Introducing new notations 

\be  
\tilde \omega_n \equiv \omega_n Z(i\omega_n), \qquad \phi_n  \equiv \phi(i \omega_n), \qquad \boldsymbol{\tilde \gamma} \equiv i \boldsymbol{\gamma} \gamma_0, \label{SMEq:tildegamma}
\ee 
one can rewrite the electron Green's function as

\be \label{SMEq:G}
\hat G(i\omega_n,\bk) = -\frac{1}{\tilde\omega_n^2 + v_F^2 k^2 + \phi_n^2} \left( \begin{array}{cc} i\tilde \omega_n + v_F (\bk \cdot \boldsymbol{\tilde \gamma}) & \phi_n \\ \phi_n &  i\tilde \omega_n - v_F (\bk \cdot \boldsymbol{\tilde \gamma}) \end{array}  \right).
\ee

The Eliashberg equation for the electron self-energy at zero momentum then takes the form 

\begin{align}
\hat \Sigma (i\omega_n ) &= \lambda^2 T\sum_m \int\frac{d^3 {\bk}}{(2\pi)^3} \left(   \begin{array}{cc}   -\gamma_i   (i\tilde \omega_m + v_F \bk \cdot \boldsymbol{\tilde \gamma}) \gamma_j & \gamma_i \gamma_j \phi_m \\ \gamma_i \gamma_j \phi_m & -\gamma_i (i\tilde \omega_m - v_F \bk \cdot \boldsymbol{\tilde \gamma}) \gamma_j        \end{array}\right)\cdot \frac{D^{ij}[i(\omega_m - \omega_n), \bk ]}{\tilde \omega_m^2 + v_F^2 k^2 + \phi_m^2} = \nonumber \\ &= \frac{\lambda^2}{\pi^2} T\sum_m \int_0^{\Lambda_0} k^2 d k \left(   \begin{array}{cc}   -i\tilde \omega_m  & \phi_m \\ \phi_m & -i\tilde \omega_m  \end{array}\right)\cdot \frac{D[i(\omega_m - \omega_n), k ]}{\tilde \omega_m^2 + v_F^2 k^2 + \phi_m^2} = \left( \begin{array}{cc} i\omega_n(1-Z_n)   & \phi_n \\ \phi_n & i\omega_n(1-Z_n)      \end{array}  \right), \label{SMEq:Eliashbergselfenergy}
\end{align}
where we have used $D^{ij}(i\omega_n,\bk) = D(i\omega_n, k) P^{ij}(\bk)$ and $\gamma_i \gamma_j \langle P^{ij}(\bk) \rangle_{\hat \bk} = 2 I$.

Analogously, the Eliashberg equation for the phonon self-energy {\it before} projecting onto the transverse sector has the form

\begin{multline}
\Pi^{ij}(i\Omega_m, \bq) = - \frac{\lambda^2}2 T\sum_n \int\frac{d^3 \bk}{(2\pi)^3} \text{Tr} \gamma^i \tau_z \hat G (i\omega_n , \bk) \gamma^j \tau_z \hat G(i\omega_n + i \Omega_m, \bk + \bq) = \\ = 4 N \lambda^2 T \sum_n \int\frac{d^3 \bk}{(2\pi)^3}   \frac{\tilde \omega_n \tilde \omega_{n+m} \delta^{ij} + v_F^2\left[ k^i (k+q)^j + k^j (k+q)^i - \bk \cdot (\bk + \bq) \delta^{ij}  \right] + \phi_n \phi_{n+m} \delta^{ij}}{\left[ \tilde \omega_n^2 + v_F^2 k^2 + \phi_n^2 \right] \cdot \left[ \tilde \omega_{n+m}^2 + v_F^2 (\bk + \bq)^2 + \phi_{n + m}^2 \right]}, \label{SMEq:Piijgen}
\end{multline}
where we have also defined $\tilde \omega_{n+m} \equiv (\omega_n + \Omega_m)\cdot Z[i(\omega_n + \Omega_m)]$.

\section{IV. Analytical calculation}

To obtain analytical results, we further adopt a number of BCS-style approximations. In particular, we neglect the normal (diagonal in Nambu space) part of electron self-energy by setting $Z(i\omega_n) = 1$ and assume that the superconducting order parameter is frequency-independent, i.e., $\phi_n = \Delta = \text{const}$. Though these assumptions are not quite accurate, we verify later that they lead to qualitatively correct results by solving the system of frequency-dependent Eliashberg equations, see Sec.~V.

We emphasize, however, that we keep the dependence of the effective interaction (phonon propagator) on $\Delta$ as it plays crucial role in our theory. Furthermore, we focus on the zero-temperature limit, which implies that the summation over the Matsubara frequencies converts into the integral according to $T \sum_n \ldots \to \int (d \omega/2\pi) \ldots$.

\subsection{IV.A Phonon self-energy and the effective interaction}

To calculate phonon self-energy (the polarization operator), we introduce the Feynmann parameter $x$. Setting $Z(i\omega_n) = 1$ and $\phi_n = \Delta = \text{const}$ in Eq.~\eqref{SMEq:Piijgen}, we find 

\be
\Pi^{ij}(i\Omega, \bq, \Delta) = 4\lambda^2 N \int_0^1 dx \int\frac{d\omega d^3 \bk}{(2\pi)^4} \frac{[\omega^2 +  \Delta^2 - x(1-x) \Omega^2]\delta^{ij} + v_F^2 [2 k^i k^j -k^2 \delta^{ij} + x(1-x)q^2(\delta^{ij} - 2\hat q^i \hat q^j)]}{\left[ \omega^2 + v_F^2 k^2 + \Delta^2 + x(1-x)(\Omega^2 + v_F^2 q^2)  \right]^2}.
\ee

This expression is ultraviolet (UV) divergent, hence, it depends on the UV momentum cutoff $\Lambda_0$. We perform the hard-cutoff regularization. Depending on the order of integration, one obtains slightly different results. These differences, however, are unimportant for our purposes and do not affect the final result. 

\subsubsection{(i) Lorentz-symmetric integration}

First, we implement the ``Lorentz-symmetric'' integration treating frequency and all components of the momentum on equal footing:

\be
\Pi^{ij}(i\Omega, \bq, \Delta) = \frac{4\lambda^2 N}{v_F^3} \int_0^1 dx \int_0^{v_F \Lambda_0}\frac{d^4 P}{(2\pi)^4} \frac{[ \Delta^2 - x(1-x) \Omega^2]\delta^{ij} +  x(1-x)v_F^2q^2(\delta^{ij} - 2\hat q^i \hat q^j)}{\left[ P^2 + \Delta^2 + x(1-x)(\Omega^2 + v_F^2 q^2)  \right]^2},
\ee
where $P =(\omega, v_F \bk)$. The advantage of this scheme is that it gets rid of the quadratic UV divergence, keeping only the logarithmic dependence on $\Lambda_0$. Performing the integration over $P$ with the hard cutoff $v_F \Lambda_0$ (in spherical coordinates) and neglecting the terms of the order $\max\{v_F q, \Omega, \Delta\}/v_F\Lambda_0$ and higher, we obtain

\be 
\Pi^{ij}(i\Omega, \bq, \Delta) \approx \frac{\lambda^2 N}{2 \pi^2 v_F^3} \int_0^1 dx \left[ \Delta^2 \delta^{ij} - x(1-x) \Omega^2\delta^{ij} +  x(1-x)v_F^2q^2(\delta^{ij} - 2\hat q^i \hat q^j)\right] \ln \frac{v_F \Lambda_0 e^{-1/2}}{\sqrt{\Delta^2 + x(1-x)(\Omega^2 + v_F^2 q^2)}}.
\ee
Performing now integration over $x$ and projecting onto the transverse sector, we find that 

\be 
\Pi^{ij}(i\Omega_m, \bq, \Delta) \to \Pi(i\Omega_m, q, \Delta) P^{ij}(\bq),
\ee
with $\Pi(i\Omega_m, q, \Delta)$ given by 

\be  
\Pi(i\Omega, q, \Delta) = \tl^2 \alpha^2 \left\{  \left( 6 \Delta^2 - \Omega^2 + v_F^2 q^2  \right) \ln \frac{v_F \Lambda_0 e^{-1/2}}{\sqrt{\Omega^2 + v_F^2 q^2}}  + \Delta^2 f_1\left( \frac{\Delta^2}{\Omega^2 + v_F^2 q^2} \right) - (\Omega^2 - v_F^2 q^2)f_2\left( \frac{\Delta^2}{\Omega^2 + v_F^2 q^2} \right)     \right\}, \label{SMEq:Pifull}
\ee 
where we have also defined $\tl^2 \equiv \lambda^2 N /12\pi^2 v_F c^2$, $\alpha \equiv c/v_F $,  and

\be
f_1(t) \equiv 3 \int_0^1 dx \ln \frac1{t+x(1-x)}, \qquad f_2(t) \equiv 3 \int_0^1 dx \, x(1-x) \ln \frac1{t+x(1-x)}.
\ee
While the integrals for $f_1(t)$ and $f_2(t)$ can be calculated analytically, we do not find it useful to present the corresponding expressions here. Throughout this work, we assume that $\tl \ll 1$ and $\alpha \ll 1$.

\subsubsection{(ii) Integration over frequency first}

If we perform integration over frequency $\omega$ from $-\infty$ to $\infty$ first, and only then integrate over momentum  $\bk$ with the hard cutoff $\Lambda_0$ in spherical coordinates, i.e., take $\int d\omega d^3\bk \to \int_0^{\Lambda_0} d^3\bk \int_{-\infty}^{\infty} d\omega$, the answer for the polarization operator changes to 

\begin{multline}  \label{SMEq:Pifreqfirst}
\Pi(i\Omega, q, \Delta) = \tl^2 \alpha^2 \left\{2 v_F^2 \Lambda_0^2  - \frac{\Omega^2}3 +   \left( 6 \Delta^2 - \Omega^2 + v_F^2 q^2  \right) \ln \frac{2 v_F \Lambda_0 e^{-7/6}}{\sqrt{\Omega^2 + v_F^2 q^2}}  + \Delta^2 f_1\left( \frac{\Delta^2}{\Omega^2 + v_F^2 q^2} \right) \right. \\ \left. - (\Omega^2 - v_F^2 q^2)f_2\left( \frac{\Delta^2}{\Omega^2 + v_F^2 q^2} \right)   \right\}.
\end{multline}

\subsubsection{(iii) Integration over momentum first}

Finally, if we integrate over momentum (in spherical coordinates) with the hard cutoff $\Lambda_0$ first and only then integrate over frequency from  $-\infty$ to $\infty$, we obtain 

\begin{multline}
\Pi(i\Omega, q, \Delta) = \tl^2 \alpha^2 \left\{6 v_F^2 \Lambda_0^2  + \Omega^2 +   \left( 6 \Delta^2 - \Omega^2 + v_F^2 q^2  \right) \ln \frac{2 v_F \Lambda_0 e^{1/2}}{\sqrt{\Omega^2 + v_F^2 q^2}}  + \Delta^2 f_1\left( \frac{\Delta^2}{\Omega^2 + v_F^2 q^2} \right) \right. \\ \left. - (\Omega^2 - v_F^2 q^2)f_2\left( \frac{\Delta^2}{\Omega^2 + v_F^2 q^2} \right)   \right\}. \label{SMEq:Pimomentumfirst}
\end{multline}

We see that there are three discrepancies between the different answers. First, answers~\eqref{SMEq:Pifreqfirst} and~\eqref{SMEq:Pimomentumfirst} have terms $\sim \tl^2 \alpha^2 v_F^2 \Lambda_0^2$ that merely shift the position of the critical point and unimportant for us. Second, there are different numerical coefficients in front of $\Lambda_0$ under the logarithm. These coefficients only redefine the UV cutoff and do not play any significant role. Finally, there is also difference in terms $\sim \tl^2 \alpha^2 \Omega^2$, which are small in the limit $\tl \ll 1$, $\alpha \ll 1$ and can be completely neglected in our calculation. For definiteness, we use the answer from Eq.~\eqref{SMEq:Pifull} for the rest of our analytical calculation.

In the nonsuperconducting state, the polarization operator equals (we also call it $\Pi_n(i\Omega, q)$ in the main text)

\be  
\Pi(i\Omega, q, 0) = \tl^2 \alpha^2  \left( v_F^2 q^2  - \Omega^2 \right) \ln \frac{v_F \Lambda_0 e^{1/3}}{\sqrt{\Omega^2 + v_F^2 q^2}}.
\ee 

The inverse phonon propagator (effective interaction) at $\Delta=0$ takes form 

\be  
D^{-1}(i\Omega, q, 0) = D_0^{-1}(i\Omega, q) - \Pi(i\Omega, q, 0) =  r + \Omega^2 + c^2 q^2 -\tl^2 \alpha^2  \left( v_F^2 q^2  - \Omega^2 \right) \ln \frac{v_F \Lambda_0 e^{1/3}}{\sqrt{\Omega^2 + v_F^2 q^2}}. \label{SMEq:Veffbarepol}
\ee 
It is clear from this expression that the static inverse propagator now has a minimum at some finite momentum $Q$ given by 

\be
Q = \Lambda_0 \exp \left(-\frac1{\tl^2} - \frac16 \right). 
\ee 

Expanding the phonon propagator at small frequencies, we obtain 

\be  
D^{-1}(i\Omega, q, 0) \approx   r + \Omega^2 (1+\alpha^2) + c^2 q^2 -\tl^2 c^2 q^2  \ln \frac{\Lambda_0 e^{1/3}}{q} = r + \Omega^2 (1+\alpha^2) - \tl^2 c^2 q^2 \ln\frac{Qe^{1/2}}{q}, \label{SMEq:Vsmallfreq}
\ee 
where we have neglected the terms of the order $\sim \tl^2 \alpha^2 \Omega^2$. If we further expand the propagator to quadratic order near $q \approx Q$, we find for $|q-Q| \lesssim Q$

\be  
D^{-1}(i\Omega, q\approx Q, 0) \approx r - c^2  \tl^2 \frac{Q^2}2 + \Omega^2 (1+\alpha^2) + \tl^2 c^2 (q-Q)^2. \label{SMEq:Veffbarepolquadr}
\ee 
We see that the finite-momentum {\it ferroelectric density wave} (FDW) order preempts the original zero-momentum FE QCP. Indeed, while the original ferroelectric transition occurs at $r_{\text{FE}}=0$, the FDW transition takes place at 

\be 
r_{\text{FDW}} = c^2 \tl^2 Q^2/ 2 >0. \label{SMEq:xiFEW}
\ee

When the superconducting order sets in, the momentum that corresponds to the minimum of  $D^{-1}$ gradually decreases from $Q$ at $\Delta=0$ to 0 at $\Delta \ge \Delta_0$, as can be seen from Fig.~\ref{Fig:Veff} of the main text. This observation indicates that the SC order suppresses FDW in favor of a uniform FE order, though to prove this statement rigorously a more careful analysis within the Ginzburg-Landau formalism is required. To find $\Delta_0$, we expand the static phonon propagator to the quadratic order in $q$ at small momenta:

\be
D^{-1}(0,q \to 0, \Delta) \approx r - 6 \Delta^2 \tl^2 \alpha^2 \ln \frac{v_F \Lambda_0 e^{-1/2}}{\Delta} + q^2 c^2 \left( 1 - \tl^2 \ln \frac{v_F\Lambda_0 e^{-1}}{\Delta} \right). \label{SMEq:Dq2exp}
\ee
The coefficient in front of $q^2$ changes sign at  $\Delta_0$ given by

\be  
\Delta_0 = \Lambda_0 v_F \exp \left( - \frac1{\tl^2} - 1  \right) = v_F Q \exp\left( -5/6 \right). \label{SMEq:Delta0}
\ee

The transition into the ordered FE or FDW state is given by the zeros of the physical mass of the transverse optical phonons defined as 
\be
m^2(r,\Delta) \equiv {\min_q D^{-1}(0,q,\Delta\,, r)} = r + {\min_q D^{-1}(0,q,\Delta\,, 0)}, \label{SMEq:m^2}
\ee
see Eq.~\eqref{eq:ms} of the main text. The relation $m^2(r_m,\Delta)=0$ defines the line $r_m(\Delta)$, which indicates the structural transition at a given $\Delta$ and is the inverse of the function $\Delta_m(r)$ considered in the main text . At $\Delta \geq \Delta_0$, $r_m(\Delta)$ marks the transition into the FE state (since the minimum of $D^{-1}$ is at $q=0$) and is given by

\be  
r_m(\Delta) = 6 \Delta^2 \tl^2 \alpha^2 \ln \frac{v_F \Lambda_0  e^{-1/2}}{\Delta} = 6\left( 1+\frac{\tl^2}2 \right) \alpha^2 \Delta^2 + 6 \Delta^2 \tl^2 \alpha^2 \ln \frac{\Delta_0}{\Delta} , \qquad \Delta \geq \Delta_0. \label{SMEq:xicrdelta}
\ee 
At  $\Delta = \Delta_0$, the finite-momentum FDW minimum becomes the zero-momentum FE minimum, and we find
\be  
r_2 = r_m (\Delta = \Delta_0) = 6e^{-5/3}\left( 1+\frac{\tl^2}2 \right) c^2 Q^2 = 6\left( 1+\frac{\tl^2}2 \right) \alpha^2 \Delta_0^2. \label{SMEq:xicr}
\ee 
These results are presented in Eqs.~\eqref{Eq:Delta0} and~\eqref{Eq:r2} of the main text.

\subsection{IV.B Gap equation}

Now we are ready to consider the equation for the superconducting gap, which is merely the off-diagonal part of Eq.~\eqref{SMEq:Eliashbergselfenergy} and reads as 

\be  
\phi_n = \frac{\lambda^2}{\pi^2} T\sum_m \int_0^{\Lambda_0} dq \, q^2 \phi_m \frac{D[i(\omega_m - \omega_n),q]}{\tilde \omega_m^2 + v_F^2 q^2 + \phi_m^2}.
\ee 

Under the assumptions that the gap is frequency- and momentum-independent, $\phi_m = \Delta$, and that the diagonal (normal) part of the electron's self-energy can be neglected, $\tilde \omega_m = \omega_m$ with $Z_m = 1$, in the zero-temperature limit this equation reduces to

\be  
1= \frac{6 \tl^2 \alpha^2 v_F^3}{\pi N} \int_{-\infty}^{\infty} d\omega \int_0^{\Lambda_0} dq \, q^2  \frac{D(i \omega ,q)}{\omega^2 + v_F^2 q^2 + \Delta^2}, \label{SMEq:gapeq}
\ee 
where, again, we have defined $\tl^2 \equiv \lambda^2 N /12\pi^2 v_F c^2$ and $\alpha \equiv c/v_F $.

The self-consistent phonon  propagator $D$ in the gap equation includes the polarization operator $\Pi(i\Omega, q,\Delta)$ given by Eq.~\eqref{SMEq:Pifull}. However, for comparison, we also consider the solutions of the gap equation obtained when using the bare phonon propagator, Eq.~\eqref{SMEq:bareGreenfunctions}, and the propagator with the bare polarization operator neglecting the effect of a finite $\Delta$, Eq.~\eqref{SMEq:Veffbarepol}. We note that a nonzero solution for $\Delta$ exists in all cases considered below, but it is significantly enhanced when the FDW fluctuations are taken into account.

\subsubsection{(i) Bare phonon propagator}

We start the analysis of the gap equation with considering the bare phonon propagator, Eq.~\eqref{SMEq:bareGreenfunctions}:

\be  
D^{-1}(i\omega, q, \Delta) = D_0^{-1}(i\omega, q) = r + \omega^2 + c^2 q^2. 
\ee 
The gap equation then takes the form 

\begin{align}
    1 = \frac{6 \tl^2 \alpha^2}{\pi N} \int_{-\infty}^{\infty} d\omega \int_0^{v_F \Lambda_0} \frac{p^2 dp}{\omega^2 + p^2  + \Delta^2} \cdot \frac1{\omega^2 + \alpha^2 p^2 + r} = \nonumber \\ =\frac{6 \tl^2 \alpha^2}{N} \int_0^{v_F \Lambda_0} \frac{p^2 dp}{\sqrt{p^2 + \Delta^2}\sqrt{\alpha^2 p^2 + r}(\sqrt{p^2 + \Delta^2} + \sqrt{\alpha^2 p^2 + r} )}.
\end{align}
The dependence of $\Delta$ on $r$ is shown in Fig.~\ref{SMFig:F(omega)}a. The maximal value is reached at $r = 0$ and equals 

\be  
\Delta_{bare} = \Delta(r=0) = v_F \Lambda_0 \exp\left[ \frac{\ln(1+\alpha) - \alpha \ln2}{1-\alpha}  \right] \exp\left[  -\frac{N(1+\alpha)}{6 \tl^2 \alpha}   \right] \approx v_F \Lambda_0  \exp\left[  -\frac{N(1+\alpha)}{6 \tl^2 \alpha}   \right]. \label{SMEq:Delta1}
\ee 
The nonzero solution disappears at $r_{\text{bare}}$ given by 

\be  
r_{bare} = v_F^2 \Lambda_0^2 \exp\left[ \frac{2\ln 2\alpha - 2\alpha \ln(1+\alpha)}{1-\alpha}  \right] \exp\left[  -\frac{N(1+\alpha)}{3 \tl^2 \alpha}   \right] \approx 4 \alpha^2 v_F^2 \Lambda_0^2  \exp\left[  -\frac{N(1+\alpha)}{3 \tl^2 \alpha}   \right].
\ee

 The typical superconducting scale in this case is set by $\Delta_{bare}$ given by Eq.~\eqref{SMEq:Delta1}. In the limit $\alpha \ll 1$, $N \gg 1$ considered in this work, it is much smaller than what one obtains if the FDW fluctuations (the polarization operator) are taken into account, $\Delta_0 \sim v_F \Lambda_0 \exp(-1/\tl^2) \gg \Delta_{bare}$  (see below).

\begin{figure}
  \centering
  \includegraphics[width=1.\columnwidth]{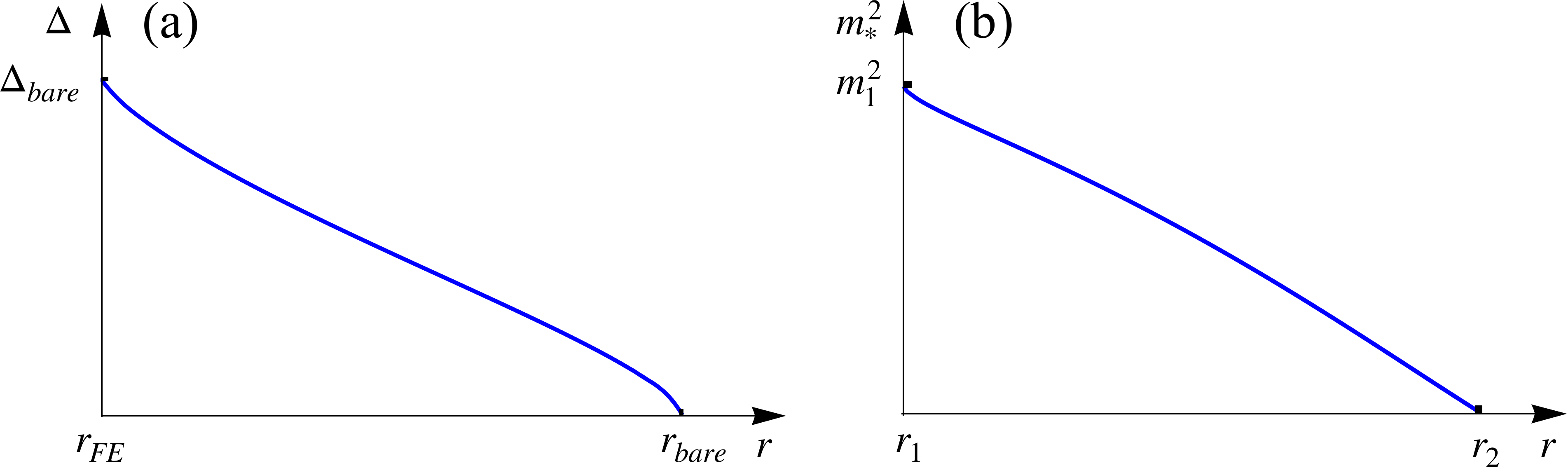}
    \caption{(a) The dependence of the nontrivial solution $\Delta$ on the tuning parameter $r$ in the case when the bare phonon propagator is used, Eq.~\eqref{SMEq:bareGreenfunctions}. It exhibits a usual shape with the maximum at $r_{\text{FE}}$ decreasing down to zero at $r_{bare}$. (b) Spectral phonon mass along the nontrivial solution $\Delta_*$, $m^2_*(r) \equiv m^2(r,\Delta_*(r))$, in the case when the feedback of $\Delta$ is taken into account. It has a maximum $m_1^2$ at $r_1 = r_{\text{FDW}} + m_1^2$, Eq.~\eqref{SMEq:m0}, and decreases to 0 at $r_2$. The inverted dependence on the bare mass $r$ explains the unusual growth of $\Delta_*$ upon increasing $r$. }
  \label{SMFig:F(omega)}
\end{figure}

\subsubsection{(ii) Phonon propagator with the bare (no feedback of $\Delta$) polarization operator}

When the bare polarization operator (constructed from the bare electron Green's functions) is taken into account, the phonon propagator is given by Eqs.~\eqref{SMEq:Veffbarepol}-\eqref{SMEq:Veffbarepolquadr}. In particular, we start with the small frequency expansion, Eq.~\eqref{SMEq:Vsmallfreq}. Performing the integration over the frequency, we find for the gap equation

\begin{align} \label{SMEq:gapeqnbarepol}
    1= \frac{6 \tl^2 \alpha^2 v_F^3}{\pi N} \int_{-\infty}^{\infty} d\omega \int_0^{\Lambda_0} \frac{q^2 dq}{\omega^2  + v_F^2 q^2 +\Delta^2} \cdot \frac1{r + \omega^2(1+\alpha^2) - \tl^2 c^2 q^2 \ln \frac{Qe^{1/2}}q} =  \\ = \frac{6 \tl^2 \alpha^2}{N} \int_0^{v_F \Lambda_0} \frac{p^2 dp}{\sqrt{p^2 + \Delta^2}\sqrt{r  - \tl^2 \alpha^2 p^2 \ln \frac{v_F Qe^{1/2}}p}}\cdot \frac1{\sqrt{(p^2 + \Delta^2)(1+\alpha^2)} + \sqrt{r  - \tl^2 \alpha^2 p^2 \ln \frac{v_F Qe^{1/2}}p}}. \nonumber
\end{align}

The dependence of $\Delta$ on $r$ is shown by the red dashed line in Fig.~\ref{Fig:Delta} of the main text. We see that $\Delta(r)$ logarithmically diverges on approaching the FDW QCP. To derive the divergence explicitly, we introduce the physical mass $m$ of the FDW phonon fluctuations according to  

\be  
m^2 \equiv r - r_{\text{FDW}} = r -  \frac{c^2 \tl^2 Q^2}2. \label{SMEq:m}
\ee 
This definition agrees with Eq.~\eqref{eq:ms} of the main text (and Eq.~\eqref{SMEq:m^2}) if one uses bare polarization operator, i.e., $D_n(0,q,r) = D(0,q,\Delta = 0, r)$ instead of $D(0,q,\Delta\,, r)$. The main contribution to the gap equation comes from the vicinity of $q\approx Q$ ($p\approx v_F Q$) and we can use the quadratic expansion from Eq.~\eqref{SMEq:Veffbarepolquadr} which is valid as long as $|q-Q| \lesssim Q$. With the logarithmic accuracy, at $m\to0$ (equivalently, at $r\to r_{\text{FDW}}$) the gap equation becomes

\be  
1 \approx \frac{6 \tl^2 \alpha^2 (v_F Q)^2}{N \sqrt{1+\alpha^2} \Delta^2} \int_{\sim -v_F Q}^{\sim v_F Q}\frac{dk}{\sqrt{m^2 + \tl^2 \alpha^2 k^2}} \approx \frac{12 \tl \alpha (v_F Q)^2}{N \sqrt{1+\alpha^2} \Delta^2} \ln \frac{v_F Q \alpha \tl}{m},
\ee 
where we have defined $k \equiv p-v_F Q$ and used that $p \approx v_F Q \ll \Delta$. It is straightforward to show that the region $v_F Q \lesssim p \lesssim v_F \Lambda_0$ only contributes a small correction to the gap equation which can be neglected to the leading order. With the logarithmic accuracy, the solution reads as 

\be 
\Delta(m) \approx v_F Q \sqrt{\frac{12 \tl \alpha}N \ln \frac{v_F Q \tl \alpha}m}
\ee 
and is applicable as long as $\Delta \gg v_F Q$, which implies the region 

\be  
m \ll \tl \alpha v_F Q \exp\left( - \frac{N}{12 \tl \alpha}  \right). \label{SMEq:smallmasympt}
\ee

This logarithmic divergence has a very clear origin. It appears due to the infrared divergence of the integral in Eq.~\eqref{SMEq:gapeqnbarepol} at $q=Q$ if $m=0$. As we show below, this divergence is cured once we take into account the self-consistent feedback of a finite $\Delta$ on the polarization operator. 

Finally, we calculate the value of $m_1$ (or $r_1)$ where the nontrivial superconducting solution vanishes. The gap equation at this point reads as 

\be  
1 = \frac{6 \tl^2 \alpha^2}{N} \int_0^{v_F \Lambda_0} \frac{p dp}{\sqrt{r_1  - \tl^2 \alpha^2 p^2 \ln \frac{v_F Qe^{1/2}}p}}\cdot \frac1{p\sqrt{1+\alpha^2} + \sqrt{r_1  - \tl^2 \alpha^2 p^2 \ln \frac{v_F Qe^{1/2}}p}}.
\ee 

It is convenient to separate the integral into two regions: $p\sim v_F Q$ and $v_F Q \lesssim p \lesssim v_F \Lambda_0$. In the first region, we use again the quadratic expansion given by Eq.~\eqref{SMEq:Veffbarepolquadr}, whereas in the second region we exploit the inequalities  $r_1 \approx r_{\text{FDW}} \lesssim  \tl^2 \alpha^2 p^2 \ln(p/v_F Q) \lesssim p^2(1+\alpha^2)$. As a result, we obtain 

\begin{align} \label{SMEq:gapbarepol}
 1 \approx    \frac{6 \tl^2 \alpha^2}{N} &\int_0^{\sim v_F Q} \frac{p dp}{\sqrt{m_1^2 + \tl^2 \alpha^2 (p-v_F Q)^2}}\cdot \frac1{p\sqrt{1+\alpha^2} + \sqrt{m_1^2 + \tl^2 \alpha^2 (p-v_F Q)^2}} +  \\ + \frac{6 \tl^2 \alpha^2}{N} &\int_{\sim v_F Q}^{v_F \Lambda_0} \frac{p dp}{\sqrt{r_1  + \tl^2 \alpha^2 p^2 \ln \frac{p}{v_F Qe^{1/2}}}}\cdot \frac1{p\sqrt{1+\alpha^2} + \sqrt{r_1  + \tl^2 \alpha^2 p^2 \ln \frac{p}{v_F Qe^{1/2}}}} \approx \nonumber \\ \approx  \frac{12 \tl^2 \alpha^2}{N \sqrt{1+\alpha^2}} &\int_0^{\sim v_F Q}\frac{dk}{\sqrt{m_1^2 + \tl^2 \alpha^2 k^2}} + \frac{6 \tl^2 \alpha^2}{N\sqrt{1+\alpha^2}} \int_{\sim v_F Q}^{v_F \Lambda_0} \frac{dp}{\tl \alpha p \sqrt{\ln(p/v_F Q)}} \approx \frac{12 \tl \alpha}{N \sqrt{1+\alpha^2}} \ln \frac{\tl \alpha v_F Q}{m_1} + \frac {12 \alpha}{N\sqrt{1+\alpha^2}}, \nonumber 
\end{align}
where, again, we have defined $m_1^2 \equiv r_1-r_{\text{FDW}}$, $k = p - v_F Q$, and used that $\ln(\Lambda_0/Q) \approx 1/\tl^2$. From this equation, we reproduce Eq.~\eqref{Eq:deltar} of the main text (with $m_1 \equiv \sqrt{\delta r}$):

\be  
m_1 \equiv \sqrt{r_1 - r_{\text{FDW}}} \sim \tl \alpha v_F Q \cdot \exp\left( -\frac{N\sqrt{1+\alpha^2}}{12 \tl \alpha} + \frac1{\tl} \right). \label{SMEq:m0}
\ee
We note that the leading exponent in this expression coincides with the one appearing in Eq.~\eqref{SMEq:smallmasympt}, where we have neglected the contribution from the region $v_F Q \lesssim p \lesssim v_F \Lambda_0$ and a factor $\sqrt{1 + \alpha^2}$ for simplicity.

\subsubsection{(iii) Phonon propagator with the self-consistent polarization operator}

We now analyze the gap equation in the case when the polarization operator self-consistently takes into account finite SC order $\Delta$  and is given by Eq.~\eqref{SMEq:Pifull}.  The nontrivial SC solution $\Delta_*(r)$ then drastically changes, as can be seen from Fig.~\ref{Fig:Delta} of the main text. It also starts at $m_1$ (equivalently, $r_1$) given by Eq.~\eqref{SMEq:m0}, however, there is no longer a nonzero solution in between $r_{\text{FDW}}$ and $r_1$. Instead, the ``dome'' extends to higher values of $r$ away from the putative FDW QCP until it reaches $\Delta_0$, Eq.~\eqref{SMEq:Delta0}, at $r_2$ given by Eq.~\eqref{SMEq:xicr}. At $r \ge r_2$, $\Delta_*(r)$ abruptly drops from $\Delta_0$ down to 0.

The absence of a solution $\D_*$ greater than $\D_0$ can be readily understood by considering the right-hand side of the gap equation~\eqref{SMEq:gapeq} in the extreme case when $\Delta=\Delta_0$ and $r=r_2$, i.e., at the point where the phonon fluctuations become gapless. The minimum of the inverse phonon propagator $D^{-1}(0,q,\Delta)$ at $\Delta=\Delta_0$ is located at $q=0$ as opposed to some finite momentum $q<Q$ at $\Delta < \Delta_0$. At this point, it has form 

\be  
D^{-1}(i\Omega,q,\Delta_0) \approx (1+\alpha^2) \Omega^2 + \tl^2 \alpha^2 \Delta_0^2 G\left( \frac{v_F q}{\Delta_0}  \right), \label{SMEq:veffsmallomega}
\ee
with 

\be  
G(x) = -\frac43(3+x^2) + \frac{(4+x^2)^{3/2} \text{ArcTanh} \left( \frac{x}{\sqrt{4+x^2}} \right)}x \approx \left\{  \begin{array}{cc} x^4/20, & x\ll 1, \\ x^2\ln x, & x\gg 1.   \end{array}     \right. \label{SMEq:F}
\ee 

The right-hand side of the gap equation at this point can be easily estimated as

\be  
\frac{6 \tl^2 \alpha^2 v_F^3}{\pi N} \int_{-\infty}^{\infty} d\omega \int_0^{\Lambda_0}q^2 dq \frac{D(i\omega, q, \Delta_0)}{\omega^2 + v_F^2 q^2 + \Delta_0^2} \approx \frac{6 \tl \alpha}N  \int_0^{v_F \Lambda_0/\Delta_0} \frac{x^2 dx}{x^2 +1} \cdot \frac1{\sqrt{G(x)}} \approx \frac{12 \alpha}{N} \ll 1.
\ee 
In this derivation, we have neglected term $\tl \alpha \sqrt{G(x)}$ compared to $\sqrt{x^2 + 1}$ after integrating over frequencies and exploited the fact that at weak coupling, $\tl \ll 1$, the main contribution to the integral comes from the region $x \gg 1$.

Consequently, we conclude that there is no a nonzero solution for the gap equation at $\Delta=\Delta_0$ in the paraelectric phase as long as $12 \alpha < N$. Furthermore, it is straightforward to check that the same is true for $\Delta > \Delta_0$. On the other hand, at $\Delta < \Delta_0$, the minimum of $D^{-1}$ is located at some finite $q$, implying that the right-hand side of the gap equation is infrared divergent when phonon mass $m$ equals 0, i.e., at $r=r_m(\Delta)$ defined below Eq.~\eqref{SMEq:m^2}. At somewhat bigger $m$, the integral becomes convergent and a self-consistent solution  $\Delta_*(r) < \Delta_0$ can be found at some finite (though exponentially small) $m$. 

Now we calculate the asymptotic behavior of the nontrivial solution as $\Delta_*(r) \to 0$ at $r\to r_1$. To that end, we expand the phonon propagator to the quadratic order in $\Delta$ and find:

\be  
D^{-1}(i\Omega, q, \Delta) \approx D^{-1}(i\Omega, q, 0) - 6\alpha^2 \tl^2 \Delta^2 \left( \ln \frac{v_F\Lambda_0}{\sqrt{\Omega^2 + v_F^2 q^2}} + \frac{\Omega^2}{\Omega^2 + v_F^2 q^2} \right) \approx D^{-1}(i\Omega, q, 0) - 6\alpha^2 \Delta^2,
\ee 
where we have used that the typical momenta are of the order $q \sim Q$. The effect of a finite $\Delta$ is merely to renormalize the bare tuning parameter as $r \to r - 6 \alpha^2 \Delta^2$. 
The solution of the gap equation is then absolutely identical to that of Eq.~\eqref{SMEq:gapbarepol} with the only difference that the number $m_1^2 = r_1 - r_{\text{FDW}}$ now should be replaced with $r - r_{\text{FDW}} - 6 \alpha^2 \Delta^2_*(r)$. This immediately leads to the answer

\be 
\Delta_*(r) \approx \sqrt{\frac{r - r_{\text{FDW}} - m_1^2}{6 \alpha^2}} = \sqrt{\frac{r - r_1}{6 \alpha^2}}, \qquad \Delta_* \ll \Delta_0. 
\ee 

To study the stability of this solution, we introduce function $f[\Delta]$ that represents the right-hand side of the gap equation (at a fixed $r$): 

\be  
f[\Delta] \equiv \frac{6 \tl^2 \alpha^2 v_F^3}{\pi N} \int_{-\infty}^{\infty} d\omega \int_0^{\Lambda_0} dq \, q^2  \frac{D(i \omega ,q)}{\omega^2 + v_F^2 q^2 + \Delta^2},
\ee 
such that the gap equation itself reads as $f[\Delta_*] = 1$. This function is related to the derivative of the free energy $F[\Delta]$ as $\partial F[\Delta]/\partial \Delta = \Delta - \Delta f[\Delta]$. Consequently, at $\Delta = \Delta_*$, one finds 

\be 
\left. \frac{\partial^2 F[\Delta]}{\partial \Delta^2} \right \vert_{\Delta = \Delta_*}= \left. - \Delta_* \frac{\partial f[\Delta]}{\partial \Delta} \right \vert_{\Delta = \Delta_*},
\ee
i.e., the sign of $f'[\Delta_*]$ determines the stability of the solution. 

At $\Delta \to 0$ and $r\to r_1$, $f[\Delta]$ is given by the right-hand side of Eq.~\eqref{SMEq:gapbarepol} with $m_1 \to \sqrt{r - r_{\text{FDW}} - 6 \alpha^2 \Delta^2}$: 

\be 
f[\Delta] \approx \frac{12 \tl \alpha}{N \sqrt{1+\alpha^2}} \ln \frac{\tl \alpha v_F Q}{\sqrt{r - r_{\text{FDW}} - 6 \alpha^2 \Delta^2}} + \frac {12 \alpha}{N\sqrt{1+\alpha^2}}, \qquad \Delta \ll \Delta_0, \, r \to r_1.
\ee
At $\Delta = \Delta_*$, we find 

\be  
\left. \frac{\partial f[\Delta]}{\partial \Delta} \right \vert_{\Delta = \Delta_*} \approx \frac{72 \tl \alpha^3 \Delta_*}{N \sqrt{1 + \alpha^2} m_1^2} > 0,
\ee 
i.e., the nontrivial solution is {\it unstable}.

To derive the asymptotic behavior of $\Delta_*(r)$ as $\Delta_* \to \Delta_0$ at $r \to r_2$, we use the low-momentum and low-frequency expansion of the bosonic propagator up to the orders $\mathcal{O}(q^4)$ and $\mathcal{O}(\Omega^2)$: 

\be
D^{-1}(i\Omega, q, \Delta) \approx m^2(r,\Delta) + (1+\alpha^2) \Omega^2 + \frac{\alpha^2 \tl^2 v_F^4}{20 \Delta^2} (q^2 - \tilde q^2)^2, \label{SMEq:Veffcrit}
\ee
with 

\be  
m^2(r,\Delta) \approx r - r_2 \left( \frac{\Delta}{\Delta_0}\right)^2 - 6 \alpha^2 \tl^2 \Delta^2 \ln\frac{\Delta_0}{\Delta} - 5\alpha^2 \tl^2 \Delta^2 \ln^2\frac{\Delta_0}{\Delta}, \qquad \tilde q^2 = \frac{10 \Delta^2}{v_F^2} \ln \frac{\Delta_0}{\Delta}, \label{SMEq:m^2exp}
\ee
where $m^2(r, \Delta)$ was calculated according to Eq.~\eqref{SMEq:m^2}. Importantly, $\tilde q^2 > 0$ for $\Delta < \Delta_0$, i.e., the inverse propagator has a minimum at a finite momentum. On the other hand, if $\Delta \geq \Delta_0$, the minimum is at $q=0$, and we easily reproduce Eq.~\eqref{SMEq:Dq2exp} with all the consequences. 

In the limit $\Delta \to \Delta_0$, $r \to r_2$ (which implies that $m\to 0$), the main contribution to the gap equation comes from momenta $|q - \tilde q| \lesssim \tilde q$, and with the leading logarithmic accuracy it can be written as 

\be 
1 \approx \frac{6 \tl^2 \alpha^2 v_F^3}{N \Delta_0^2 \sqrt{1 + \alpha^2}} \int_0^{\sim \tilde q} \frac{q^2 dq}{\sqrt{m^2_* + \frac{\alpha^2 \tl^2 v_F^4}{20 \Delta_0^2}(q^2 - \tilde q^2)^2}} \approx \frac{12\sqrt{5}\alpha \tl v_F \tilde q}{N \Delta_0 \sqrt{1 + \alpha^2}} \ln \frac{\alpha \tl v_F^2 \tilde q^2}{m_* \Delta_0}, \label{SMEq:gapsolutionneardelta0}
\ee 
where we have defined $m_*(r)\equiv m(r,\Delta_*(r))$. Expanding to the leading order in $\delta = \Delta_0 - \Delta_* \ll \Delta_0$, we obtain

\be  
1 \approx \frac{60\sqrt{2}\alpha \tl}{N\sqrt{1 + \alpha^2}} \sqrt{\frac{\delta}{\Delta_0}} \ln \frac{\alpha \tl \delta}{m_*} \qquad \Longrightarrow \qquad m_*\sim \alpha \tl \delta \exp\left( - \frac{N\sqrt{1 + \alpha^2}}{60\sqrt{2} \alpha \tl} \sqrt{\frac{\Delta_0}{\delta}}   \right). \label{SMEq:m(delta)}
\ee
In the above derivation, we have assumed that $10 \delta \lesssim \Delta_0$ and $(v_F \tilde q/\Delta_0) \ln(\alpha \tl v_F^2 \tilde q^2/m_* \Delta_0)\gg 1$, which is obviously satisfied in the limit $\alpha\ll1$, $\tl \ll 1$.

Equation~\eqref{SMEq:m(delta)} shows that the phonon spectral mass $m(r, \Delta_*(r))$ is exponentially small at the nontrivial superconducting solution when $\Delta_* \to \Delta_0$. In fact, Eq.~\eqref{SMEq:m0} demonstrates that it remains exponentially small even at $\Delta_*=0$. This result implies that the nontrivial solution $\Delta_*(r)$ with the exponential accuracy coincides with the curve $\Delta_m(r)$ defined as $m^2(r,\Delta_m)=0$, or with its inverse $r_m(\Delta)$. From Eq.~\eqref{SMEq:m^2exp}, we obtain 

\be 
r_m(\Delta) \approx  r_2 \left( \frac{\Delta}{\Delta_0}\right)^2 + 6 \alpha^2 \tl^2 \Delta^2 \ln\frac{\Delta_0}{\Delta} + 5\alpha^2 \tl^2 \Delta^2 \ln^2\frac{\Delta_0}{\Delta}, \qquad \Delta < \Delta_0. \label{SMEq:rmDelta<Delta0}
\ee 
The first two terms of this expression exactly correspond to Eq.~\eqref{SMEq:xicrdelta} derived for $\Delta \ge \Delta_0$, while the last term represents the leading $\tilde q^4 \propto \delta^2$ correction due to the fact that the minimum of $D^{-1}$ is at a finite momentum if $\Delta < \Delta_0$. Expanding further Eq.~\eqref{SMEq:rmDelta<Delta0} to the leading order in $\delta$, we find

\be  
\delta(r) = \Delta_0 - \Delta_*(r) \approx \frac12 \left( 1 + \frac{\tl^2}2\right)\frac{r_2 - r}{r_2}\Delta_0, \qquad \delta \ll \Delta_0. \label{SMEq:delta}
\ee 
Since $\delta > 0$, the solution exists only for $r<r_2$, in agreement with Fig.~\ref{Fig:Delta} from the main text.

When using the expansion~\eqref{SMEq:Veffcrit}, we dropped the terms of the form $\alpha^2 \tl^2 (v_F^2 q^2)^{n+1-l} (\Omega^2)^l/(\Delta^2)^n$, with $n \ge 1$ and $0 \le l \le n+1$ (except for the one with $l=0$, $n=1$). Since the typical momenta are of the order $q\sim \tilde q$ and typical frequencies are $\Omega \sim \alpha \tl v_F^2 \tilde q^2/\Delta_0$, these terms have additional smallness of at least $(v_F \tilde q)^2/\Delta_0^2 \sim \delta/\Delta_0$ and can be safely neglected in the limit $\delta \ll \Delta_0$.

Again, to study the stability of the solution $\Delta_*(r)$ at $\Delta \to \Delta_0$ and $r \to r_2$, we consider $f[\Delta]$, which in this limit is given by Eq.~\eqref{SMEq:gapsolutionneardelta0} with $m_*$ being substituted with $m(r,\Delta)$:

\be  
f[\Delta] \approx \frac{60 \sqrt{2} \alpha \tl}{N\sqrt{1+\alpha^2}}\sqrt{\frac{\Delta_0 - \Delta}{\Delta_0}}\ln\frac{\alpha \tl(\Delta_0 - \Delta)}{m(r,\Delta)}, \qquad m^2(r,\Delta) \approx r-r_2 + \frac{2 r_2}{1+\frac{\tl^2}{2}}\frac{\Delta_0 - \Delta}{\Delta_0}, \qquad \Delta \to \Delta_0, \, r\to r_2.
\ee 
At $\Delta = \Delta_*$, we find 

\be  
\left. \frac{\partial f[\Delta]}{\partial \Delta} \right \vert_{\Delta = \Delta_*} \approx \frac{360 \sqrt{2} \tl \alpha^3 \sqrt{\Delta_0 (\Delta_0 - \Delta_*)}}{N \sqrt{1 + \alpha^2} m_*^2} > 0,
\ee 
where $\Delta_*$ and $m_*$ are given by Eqs.~\eqref{SMEq:delta} and~\eqref{SMEq:m(delta)}, correspondingly. We see again that the nontrivial solution is unstable, in agreement with the discussion from the main text.

The instability of the nontrivial solution also reveals itself in the unusual behavior of the spectral phonon mass $m$ along this solution. We show the dependence of $m_*^2(r) \equiv m^2(r, \Delta_*(r))$ on the tuning parameter $r$ in Fig.~\ref{SMFig:F(omega)}b. We see that as we tune the system away from the critical point, i.e., increase the bare mass $r$, the spectral mass $m_*$ decreases. It can be easily shown that this line starts from the value $m_1^2 = r_1 - r_{\text{FDW}}$ given by Eq.~\eqref{SMEq:m0} at $r = r_1$ and decreases down to 0 at $r = r_2$, see Eq.~\eqref{SMEq:xicr}. Such inverted dependence originates from the nontrivial feedback effect of the superconducting gap $\Delta$ on the phonon propagator discussed in the main text and explains the growth of $\Delta_*(r)$ upon increasing $r$. Indeed, larger $r$ correspond to smaller spectral mass, which implies stronger FDW fluctuations leading to higher values of $\Delta_*$.

\subsection{IV.C Superconducting stiffness }

We conclude the analytical section by calculating the superconducting stiffness. The calculation is most easily performed in the paraelectric phase, while it can be shown that the result remains qualitatively the same in the ferroelectric phase and we expect it to hold in the FDW phase as well.

In the presence of a nonzero vector potential $\bA$, the electron's Hamiltonian in Nambu space is obtained by the substitution $\bk \to \bk - e\bA \tau_z$ (we set the speed of light equal to 1 here). It implies that the correction to the electron Green's function, Eq.~\eqref{SMEq:G}, is given by 

\be  
\delta \hat G^{-1} = e v_F \boldsymbol{\tilde \gamma} \cdot \bA,
\ee 
while the current operator equals

\be  
\hat \bj \equiv \frac{\delta \hat G^{-1}}{\delta \bA} = e v_F \boldsymbol{\tilde \gamma},
\ee 
where $\boldsymbol{\tilde \gamma}$ is defined in Eqs.~\eqref{SMEq:tildegamma} and~\eqref{SMEq:gammadef}. The superconducting stiffness $\rho_s^{\alpha \beta}$ defined through the relation $j^{\alpha} = \rho_s^{\alpha \beta} A_\beta$ is then given by the Kubo formula and to the leading logarithmical order equals

\be 
\rho_s^{\alpha \beta} = - \text{Tr}[\hat j^\alpha \hat G \hat j^\beta \hat G] = -\frac{2Ne^2}{\pi^2 v_F } \Delta^2 \ln \frac{v_F \Lambda_0}{\Delta} \delta^{\alpha \beta}.
\ee

\section{V. Numerical Solution of the Eliashberg equations}

To understand how our analytical  results are affected by the approximations we have used, we supplement them with a fully self-consistent solution of the Eliashberg equations. This analysis includes both the frequency-dependent order parameter $\phi(i\w)$ and the normal part of the electron's self-energy encapsulated in a quasiparticle weight $Z(i\w)$. It is important to note that we still neglect vertex corrections and the momentum dependence of the self-energy.

\subsection{V.A Setup for calculating the polarization operator}

Before jumping to numerical integration, we simplify the task by performing the momentum integration for the polarization operator analytically. Under the assumptions that the superconducting order parameter has $s$-wave symmetry and that electron's self-energy is momentum-independent, the polarization operator before projecting onto the transverse sector is given by Eq.~\eqref{SMEq:Piijgen}. Introducing Feynman parameter $x$, it can be rewritten as 

\be
\Pi^{ij}(i\Omega_m, \bq) =  2 N \lambda^2 T\sum_n \int_0^1 dx \int\frac{d^3 p}{(2\pi)^3} \frac{2(\tilde \omega_n \tilde \omega_{n+m} + \phi_n \phi_{n+m} - \frac13 p^2 v_F^2)\delta^{ij} + 2v_F^2 x(1-x)(q^2 \delta^{ij} - 2 q^i q^j) }{(v_F^2 p^2 + R^2)^2},
\ee
where we have shifted the momentum variable as $\bk = \bp -(1- x) \bq$ and defined

\be  
R^2 = v_F^2 q^2 x(1-x) + x \tilde \omega_{n+m}^2 + (1-x) \tilde \omega_n^2 + x \phi_{n+m}^2 + (1-x) \phi_n^2.
\ee

The momentum integration can be performed exactly using the dimensional regularization with the identities

\be  
\int\frac{d^3 p}{(2\pi)^3}  \frac1{\left( v_F^2 p^2 + R^2 \right)^2} = \frac1{8\pi v_F^3 R}, \qquad \int\frac{d^3 p}{(2\pi)^3}  \frac{p^2}{\left( v_F^2 p^2 + R^2 \right)^2} = -\frac{3R}{8\pi v_F^5}.
\ee 
The result reads as 

\be
\Pi^{ij}(i\Omega_m, \bq, \phi) =  \frac{N \lambda^2}{2 \pi v_F^3} T\sum_n \int_0^1 dx  \frac{(\tilde \omega_n \tilde \omega_{n+m} + \phi_n \phi_{n+m} + R^2)\delta^{ij} + v_F^2 x(1-x)(q^2 \delta^{ij} - 2 q^i q^j) }{R}.
\ee
After projecting onto the transverse sector in the $T=0$ limit, one finds that $\Pi^{ij}(i\Omega_m, \bq, \phi) \to \Pi(i\Omega, q, \phi) P^{ij}(\bq)$, with

\be
\Pi(i\Omega, \bq, \phi) =  3 \tl^2 \alpha^2 \int_{-\vf \Lambda_0}^{\vf \Lambda_0} d\omega \int_0^1 dx  \frac{\omega (\omega + \Omega) Z(\omega) Z(\omega + \Omega) + \phi(\omega) \phi(\omega + \Omega) + x(1-x) v_F^2 q^2 + R^2}{R},
\ee
and $R^2$ at $T=0$ takes form

\be  
R^2 = x(1-x) v_F^2 q^2 + x(\omega + \Omega)^2 Z^2(\omega + \Omega) + (1-x) \omega^2 Z^2(\omega) + x \phi^2(\omega + \Omega) + (1-x) \phi^2(\omega).
\ee 
If the frequency dependence of the gap function and the quasiparticle weight renormalization are neglected, $\phi(\omega) = \Delta$ and $Z(\omega) = 1$, we reproduce Eq.~\eqref{SMEq:Pimomentumfirst}. Since this result exhibits quadratic UV divergence $\propto \Lambda_0^2$, it is convenient to subtract $\Pi(0,0,0)$, which is merely a constant, and consider instead

\begin{multline}
\tilde \Pi(i\Omega, q, \phi) \equiv \Pi(i\Omega, q, \phi) - \Pi(0,0,0) = \\ = 3 \tl^2 \alpha^2 \int_{-\vf \Lambda_0}^{\vf \Lambda_0} d\omega \int_0^1 dx \left\{ \frac{\omega (\omega + \Omega) Z(\omega) Z(\omega + \Omega) + \phi(\omega) \phi(\omega + \Omega) + x(1-x) v_F^2 q^2 + R^2}{R} - 2\left|\omega Z(\omega)\right|  \right\}.
\end{multline}

To simplify the numerical evaluation even further, we perform integration over $x$ explicitly. It is convenient to introduce 

\be  
J_0\equiv \int_0^1 \frac{dx}R = \frac1{v_F q} \left( \arctan \frac{v_F^2 q^2 + (\widetilde{\omega + \Omega})^2 - \tilde \omega^2 + \phi^2_{\omega + \Omega} - \phi^2_{\omega}}{2 v_F q \sqrt{\tilde \omega^2 + \phi^2_{\omega}}} + \arctan \frac{v_F^2 q^2 - (\widetilde{\omega + \Omega})^2 + \tilde \omega^2 - \phi^2_{\omega + \Omega} + \phi^2_{\omega}}{2 v_F q \sqrt{(\widetilde{\omega+\Omega})^2 + \phi^2_{\omega + \Omega}}} \right),
\ee 
where we have defined $\tilde \omega = \omega Z(\omega)$, $\widetilde{\omega+\Omega} = (\omega + \Omega)Z(\omega + \Omega)$, $\phi_{\omega} = \phi(\omega)$, and $\phi_{\omega + \Omega} = \phi(\omega + \Omega)$. Analogously, we define

\begin{align}
    J_1 &\equiv \int_0^1  \frac{x(1-x) dx}R = \frac{3\left[ (\widetilde{\omega+\Omega})^2 - \tilde \omega^2 + \phi^2_{\omega + \Omega} - \phi^2_{\omega} \right]\cdot \left[  \sqrt{(\widetilde{\omega+\Omega})^2 + \phi^2_{\omega + \Omega}} - \sqrt{\tilde \omega^2 + \phi^2_{\omega}}\right]}{4 v_F^4 q^4} + \nonumber\\&+ \frac{\sqrt{(\widetilde{\omega+\Omega})^2 + \phi^2_{\omega + \Omega}} + \sqrt{\tilde \omega^2 + \phi^2_{\omega}}}{4 v_F^2 q^2} - \left\{ \frac{3 \left[ (\widetilde{\omega+\Omega})^2 - \tilde \omega^2 + \phi^2_{\omega + \Omega} - \phi^2_{\omega} \right]^2   }{8 v_F^4 q^4} + \frac{(\widetilde{\omega+\Omega})^2 + \tilde \omega^2 + \phi^2_{\omega + \Omega} + \phi^2_{\omega}}{4 v_F^2 q^2} -\frac18 \right\}J_0, \nonumber \\
    J_2 &\equiv \int_0^1 R dx = \frac{\left[ (\widetilde{\omega+\Omega})^2 - \tilde \omega^2 + \phi^2_{\omega + \Omega} - \phi^2_{\omega} \right]\cdot \left[    \sqrt{\tilde \omega^2 + \phi^2_{\omega}} - \sqrt{(\widetilde{\omega+\Omega})^2 + \phi^2_{\omega + \Omega}}\right]}{4 v_F^2 q^2} + \nonumber \\ &+ \frac{\sqrt{\tilde \omega^2 + \phi^2_{\omega}} + \sqrt{(\widetilde{\omega+\Omega})^2 + \phi^2_{\omega + \Omega}}}{4} + \frac{J_0}{8 v_F^2 q^2}\left\{ v_F^4 q^4 + 2 v_F^2 q^2 \left[ (\widetilde{\omega+\Omega})^2 + \tilde \omega^2 + \phi^2_{\omega + \Omega} + \phi^2_{\omega}  \right]  + \right. \nonumber \\ &\left.+  \left[ (\widetilde{\omega+\Omega})^2 - \tilde \omega^2 + \phi^2_{\omega + \Omega} - \phi^2_{\omega}  \right]^2 \right\}. 
\end{align}
Using the relation

\begin{align}
   v_F^2 q^2 J_1 + J_2 & = \frac{\left[ (\widetilde{\omega+\Omega})^2 - \tilde \omega^2 + \phi^2_{\omega + \Omega} - \phi^2_{\omega} \right]\cdot \left[ \sqrt{(\widetilde{\omega+\Omega})^2 + \phi^2_{\omega + \Omega}}   -    \sqrt{\tilde \omega^2 + \phi^2_{\omega}} \right]}{2 v_F^2 q^2} + \\ &+ \frac{\sqrt{\tilde \omega^2 + \phi^2_{\omega}} + \sqrt{(\widetilde{\omega+\Omega})^2 + \phi^2_{\omega + \Omega}}}{2} + \frac{J_0}{4 v_F^2 q^2}\left\{ v_F^4 q^4 -  \left[ (\widetilde{\omega+\Omega})^2 - \tilde \omega^2 + \phi^2_{\omega + \Omega} - \phi^2_{\omega}  \right]^2 \right\},\nonumber
\end{align}
we find

\be  \label{eq:sup:Pi_full}
\tilde \Pi(i\Omega, q, \phi) = 3 \tl^2 \alpha^2 \int_{-v_F \Lambda_0}^{v_F \Lambda_0} d\omega \left\{ (\tilde \omega (\widetilde{\omega + \Omega}) + \phi_{\omega} \phi_{\omega + \Omega})J_0 + v_F^2 q^2 J_1 + J_2 - 2 |\tilde \omega|  \right\}.
\ee

\subsection{V.B Details and results of the numerical integration}

\begin{figure}
  \centering
  \includegraphics[width=1.\columnwidth]{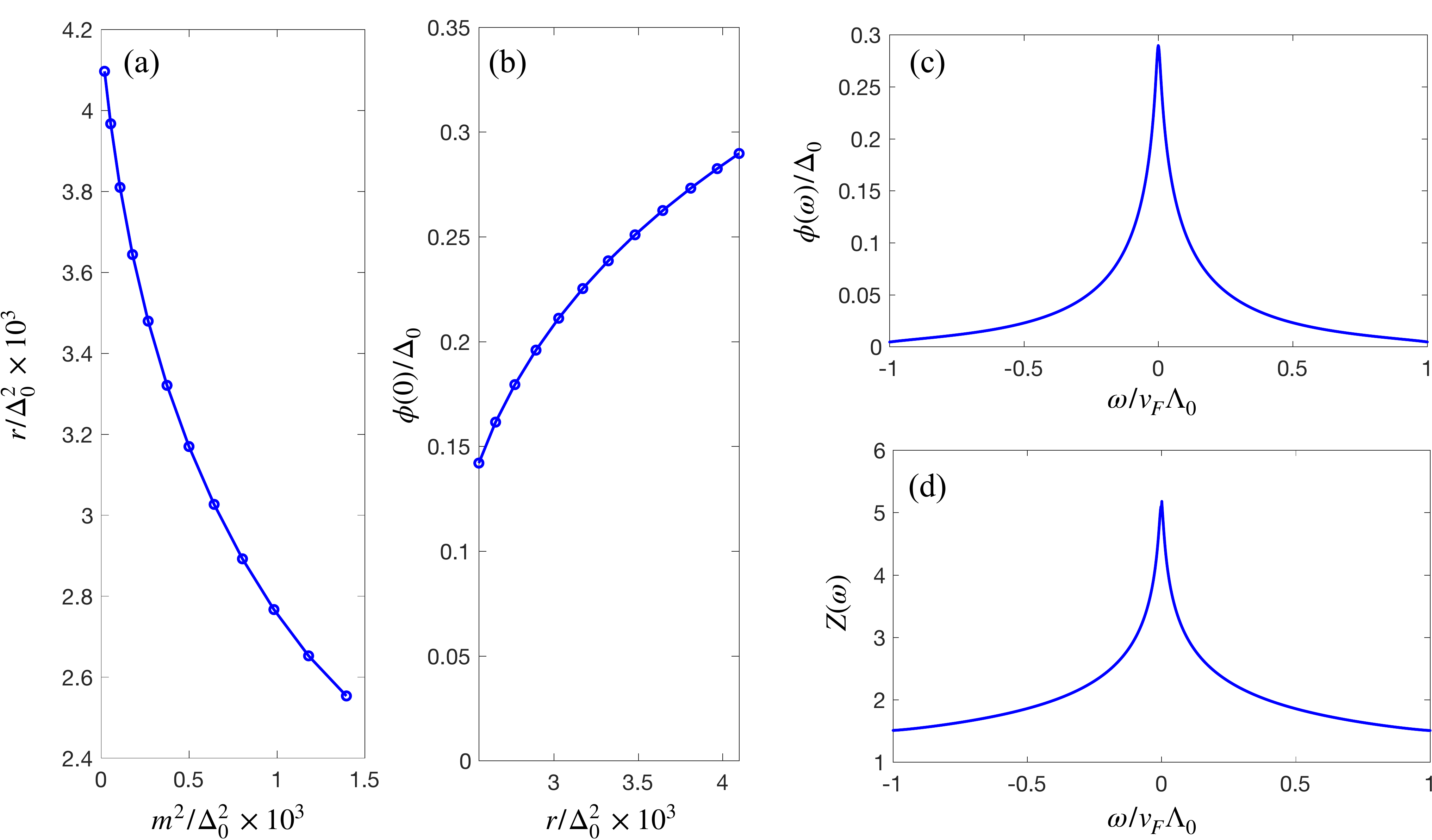}
    \caption{Results of the self-consistent solution of the coupled Eliashberg equations for $\tilde \lambda^2 = 0.35$, $\alpha = 0.2$, and $N = 1$. (a)~The tuning parameter $r$ vs. phonon spectral mass $m$. The calculation is performed at a fixed $m$ making the instability attainable within standard minimization techniques. The value of $r$ is then obtained in retrospect. (b) The peak of the order parameter $\phi(0)$ vs. the tuning parameter $r$. (c), (d) The frequency-dependent order parameter $\phi(\w)$ and quasiparticle weight $Z(\w)$ for $m/\D_0 = 0.0043$.  }
  \label{Fig:sup:numerics}
\end{figure}

To find a nontrivial self-consistent superconducting solution, we solve the system of the coupled Eliashberg equations~\eqref{eq:sup:Pi_full} and~\eqref{SMEq:Eliashbergselfenergy} iteratively. Numerical integrations are performed trapezoidaly, taking the integrals over momentum first and then over frequency. We use an equally-spaced momentum grid between $q = 0.001 Q$ and $3Q$ with $N^d_q = 60$ points and then additional $N^s_q = 60$ points between $3Q$ and $3 \Lambda_0$. The frequency on the other hand is taken to be equally spaced between $-v_F \Lambda_0$ and $v_F \Lambda_0$ with $N_\omega = 1200$ points. The results are not strongly dependent on $N_q ^{d,s}$, but they do depend strongly on the frequency spacing. In particular, it is essential that the smallest frequency $\w_{min} = 2 v_F \Lambda_0/N_\w$ is small enough. This parameter takes the role of an effective temperature and thus it must be much smaller than the gap at zero temperature.

As mentioned in the main text the solution of the gap equation~\eqref{Eq:gapsimple} is an instability point. The origin of the instability is the feedback of $\Delta$ in the polarization function, which causes the phonon spectral gap, Eq.~\eqref{eq:ms}, to become negative. 
Numerically, however, it is much more challenging to find an unstable point as compared to a stable one. To deal with this problem, we fix the phonon spectral mass $m$ to be always in the paraelectric phase and obtain the corresponding tuning parameter $r$ as a function of $\phi$, $Z$, and $m$. By doing this, we force the unstable solution to become stable in the space of fixed $m$, which enables the use of standard iterative techniques.

The iterative procedure is initiated by $Z_0(i\omega) = 1$, $\phi_0(i\w) = 0.01 \D_0$, and $\Pi(i\W,q,\phi_0(i\w))$. We find that after an initial transient of $10$ to $20$ iterations the difference function $\eta = N_\w ^{-1}\sum_\w \left(|1-\phi_{l}(i\w)/\phi_{l+1}(i\w)|+|1-Z_{l}(i\w)/Z_{l+1}(i\w)|\right)$ decays exponentially with the iteration number $l$. The exponential decay time diverges as the critical value $r_1$ is approached. We cutoff the iteration at 140 iterations, which produces $\eta < 10^{-3}$ for all the data points presented here. 

The results for $N = 1$, $\tilde \lambda^2 = 0.35$, and $\alpha= 0.2$ are presented in Fig.~\ref{Fig:sup:numerics}. Panel (a) shows the dependence of the bare mass $r$ on the spectral mass $m$. As in the case without frequency dependence, they depend inversely on each other. We also identify the unusual dependence of $\phi(0)$ on $r$ in panel (b). Putting these two behaviors together we find that the increase in $\phi$ reduces the spectral mass, just as in the case where frequency dependence is neglected. This is the key mechanism to drive the first-order transition into the paired state with ferroelectric (or FDW) order. We therefore conclude that the first-order transition is not an artifact of neglecting the frequency dependence in $\phi$ and $Z$. In panels (c) and (d) we plot the solutions for $\phi(i\w)$ and $Z(i\w)$ for $r/\D_0^2 = 0.0041$ (corresponding to $m/\D_0 = 0.0043$).

We note that the inclusion of the normal part of the self-energy (encapsulated in $Z(i\w)$) and frequency dependence suppresses $\phi(0)$  when compared with the solution of the frequency-independent  BCS-like  gap equation~\eqref{Eq:gapsimple} with the same parameters. We also find that the value of $r_1$ is shifted to smaller $r$ toward $r_{\text{FDW}}$. This implies that the value of the nontrivial solution inside the ferroelectric phase, which is bounded from below by the instability curve, is most likely diminished by the inclusion of $Z(i\omega)$.

\section{VI. Speculations about the phase diagram including the ferroelectric phase} 
In this paper we have argued that a ferroelectric quantum critical point is preempted by a first-order transition into a superconducting and ferroelectric state (uniform or finite-momentum). 
We have made this point based on calculations performed in the paraelectric phase. In particular, we have found that when the tuning parameter $r$ is in the regime $r_1<r<r_2$ the paraelectric solution with $\D = 0$ is stable (represented by the blue dots in Fig.~\ref{Fig:sup:specs}). 
However, for $\D > \D_{*}(r)$, the Ginzburg-Landau theory is unstable toward larger $\D$.  Eventually,  for $\D >  \D_m(r)$, the phonon spectral mass becomes negative and the free energy becomes unstable toward a finite average optical phonon displacement $\bs u$ (ferroelectric order). Here we discuss possible scenarios for the free energy when it is extended into the space of a finite $\bs u$. 

A schematic two-dimensional contour plot of the free energy corresponding to the regime $r_1<r<r_2$  appears in Figs.~\ref{Fig:sup:specs}(a,b). Panel (a) shows the likely scenario where a new stable solution is both ferroelectric and superconducting (red dot). 
However, within our formalism we can not rule out the scenario where the minimum in the ferroelectric phase lies on the $\D = 0$ axis (no superconductivity), as shown in panel (b).
We deem this scenario unlikely for two main reasons. The first one is that such a landscape is highly complicated and it is hard to imagine a mechanism that would generate it. Second, in this case the second minimum is on the line $\D = 0$ and can therefore be found without considering superconductivity. Notwithstanding, in Ref.~\cite{KoziiBiRuhman2019} some of us found that for $\alpha \ll 1$ the electron's contribution to the quartic $u^4$ term in the Ginzburg-Landau free energy is positive. Thus, if there is a second minimum it can only be seen at the order of $u^6$.  

Finally, in panel~(c) we plot the likely scenario in the regime $r_{\text{FDW}}<r<r_1$, where the solution $\D = 0$ is no longer stable and there is a flow toward finite $\D$ and $u$. Also here, we can not rule out a curved flow going back to the $\D = 0$ line, but find it highly unlikely.

\begin{figure}
  \centering
  \includegraphics[width=0.75\columnwidth]{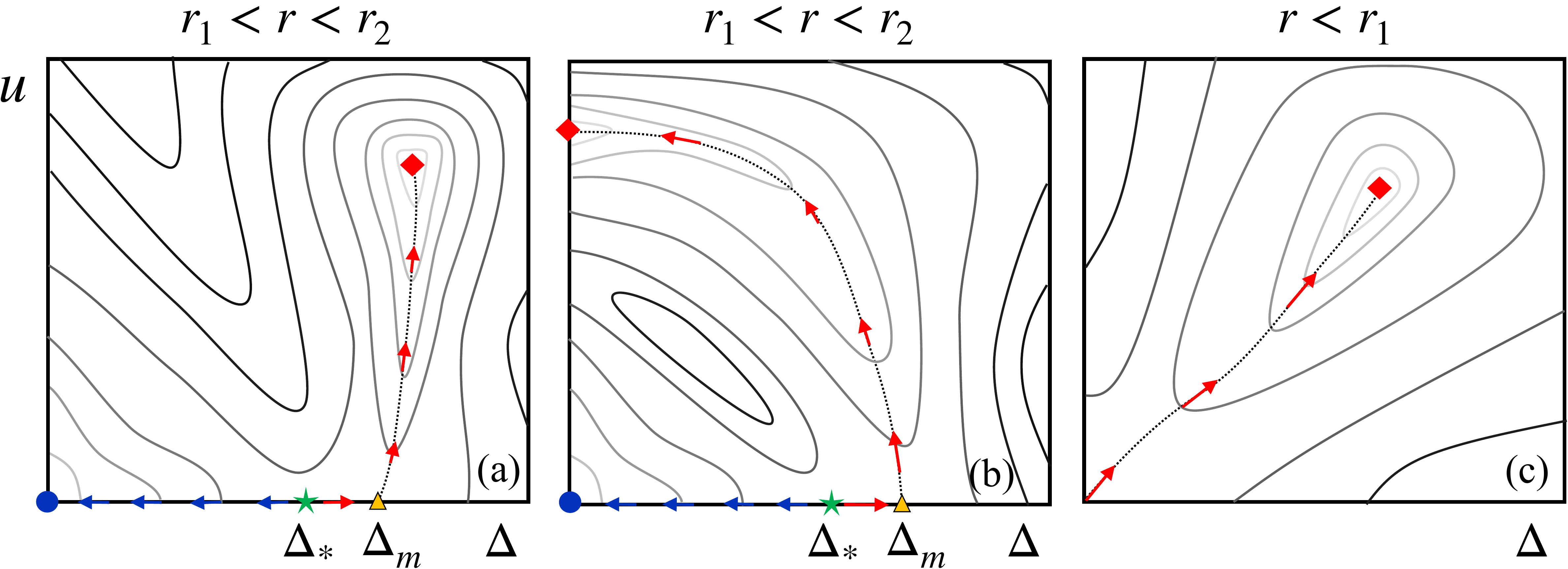}
    \caption{ Schematic contour plots of the free energy in the space of the pairing gap $\D$ and the ferroelectric order parameter $u$. (a)~The likely scenario for $r_1<r<r_2$, where the point $\D = 0$ and $u = 0$ (blue dot) is stable. The point $\D_*$ marks the drainage divide point beyond which the flow is toward larger $\D$. At $\Delta = \D_m$, the phonon spectral mass becomes negative indicating the instability toward finite $u$. Eventually there is a second stable point where both $\D$ and $u$ are finite (red dot). (b)~An extreme scenario that can not be ruled out within our formalism. 
    The lowest energy path between the two stable points (blue circle and red diamond) is through a finite $\D$, while the points themselves are at $\D = 0$. This scenario is unlikely both because it is a highly complicated trajectory and also because we find no indications of a first-order transition at $\D = 0$. (c)~In the regime $r<r_1$ the point $\D = 0$ and $u=0$ ceases to be stable and the only stable solution is at finite $\D$ and $u$.  }
  \label{Fig:sup:specs}
\end{figure}

\section{VII. An Estimate for the dimensionless coupling constant $\tilde\lambda$}
In this paper, we found that  the phase diagram is controlled by an emergent scale $Q \sim \exp\left[{-{1/ \tilde \lambda^2}}\right]$.
Therefore, it is constructive to  estimate the  value of the dimensionless parameter $\tilde\lambda^2$ in a real material. For PbTe/SnTe, we find
\begin{align}\label{SMEq:exp_lambda_estimate}
   \tilde \lambda^2 = {N \lambda^2 \over 12\pi^2 v_F c^2} \approx 0.27.
\end{align}
In this estimate, we used $N=4$, $v_F = 10^6\,\mathrm{m/s}$~\cite{Assaf2016}, and $c = 3\times 10^3$ m/s~\cite{Jacobsen_2013,AN2008417}.  The value of the bare coupling $\lambda$ in Eq.~\eqref{Eq:S} is estimated from the displacement of the Dirac points in momentum space, $k_0$, relative to the high symmetry point $L$ due to the static distortion, $x_0$, which is observed in ferroelectric SnTe. This momentum shift is found to be $k_0 \approx 0.15\,\mathrm{\AA}^{-1}$~\cite{tanaka2012experimental}, while the atomic displacement is smaller than $x_0 = 0.2\, \mathrm{\AA}$~\cite{yang2018non,wang2018two}. However, before plugging these values in Eq.~\eqref{SMEq:exp_lambda_estimate} we must also convert the field $\varphi$ in Eq.~\eqref{Eq:S} to a proper displacement field $x=\varphi \sqrt{\hbar/\rho_m}$, where $\rho_m\approx 8.1$ g cm$^{-3}$ is the mass density of PbTe (SnTe has $\rho_m=6.18$ g cm$^{-3}$, leading to even bigger $\tilde \lambda$). In this way, $x$ has units of length. Then, the coupling constant is given by 
\be
\lambda = {v_F k_0 \sqrt{\hbar} \over  x_0 \sqrt{\rho_m}} \approx 8.6\times 10^{6}\,(\mathrm{m/s})^{3/2}\,.
\ee
%\[ \lambda = {v_F k_0 \sqrt{\hbar} \over  x_0 \sqrt{\rho_m}} \approx 8.7\times 10^{-11}\,\mathrm{ J^2 m /kg }\,.\]
Putting all these numbers together we arrive at Eq.~\eqref{SMEq:exp_lambda_estimate}. While this is just a rough estimate, we see that this coupling to optical phonons can be significant, thus justifying the potential relevance of our study to real materials.  

%It should be noted however, that the real material brings many additional complications such as crystal symmetry breaking, which were neglected in this paper. Therefore, the above estimate only comes to show that the coupling to  such phonons can be important. 

\end{widetext}

\end{document}